\documentclass[final,5p,times,twocolumn,authoryear]{elsarticle}

\usepackage{graphicx}
\usepackage{amssymb}
\usepackage{natbib}
\usepackage{hyperref}

\journal{Icarus}

\begin{document}

\begin{frontmatter}

\title{A study of 3-dimensional shapes of asteroid families with an application to Eos}

\author{M.~Bro\v z}

\address{Institute of Astronomy, Charles University, Prague, V Hole\v sovi\v ck\'ach 2, 18000 Prague 8, Czech Republic e-mail: mira@sirrah.troja.mff.cuni.cz}

\author{A.~Morbidelli}

\address{Observatoire de la C\^ote d'Azur, BP 4229, 06304 Nice Cedex 4, France, e-mail: morby@oca.eu}


\begin{abstract}
In order to fully understand the shapes of asteroids families
in the 3-dimensional space of the proper elements
$(a_{\rm p}, e_{\rm p}, \sin I_{\rm p})$ it is necessary
to compare observed asteroids with N-body simulations.
To this point, we describe a rigorous yet simple method which
allows for a selection of the observed asteroids,
assures the same size-frequency distribution of synthetic asteroids,
accounts for a background population,
and computes a~$\chi^2$~metric.
We study the Eos family as an example,
and we are able to fully explain its non-isotropic features,
including the distribution of pole latitudes~$\beta$.
We confirm its age $t = (1.3\pm0.3)\,{\rm Gyr}$;
while this value still scales with the bulk density,
it is verified by a Monte-Carlo collisional model.
The method can be applied to other populous families (Flora, Eunomia, Hygiea , Koronis, Themis, Vesta, etc.).
\end{abstract}

\end{frontmatter}



\section{Introduction}\label{sec:introduction}

\begin{figure}
\centering
\includegraphics[height=6cm]{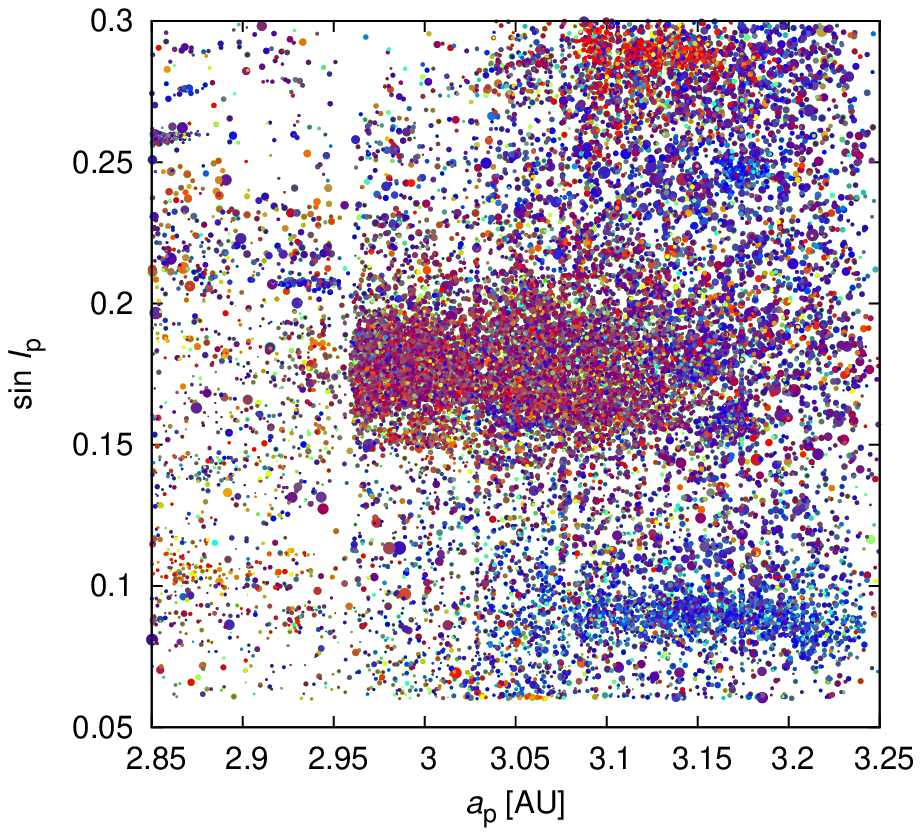}
\vskip-.3cm
\includegraphics[height=6cm]{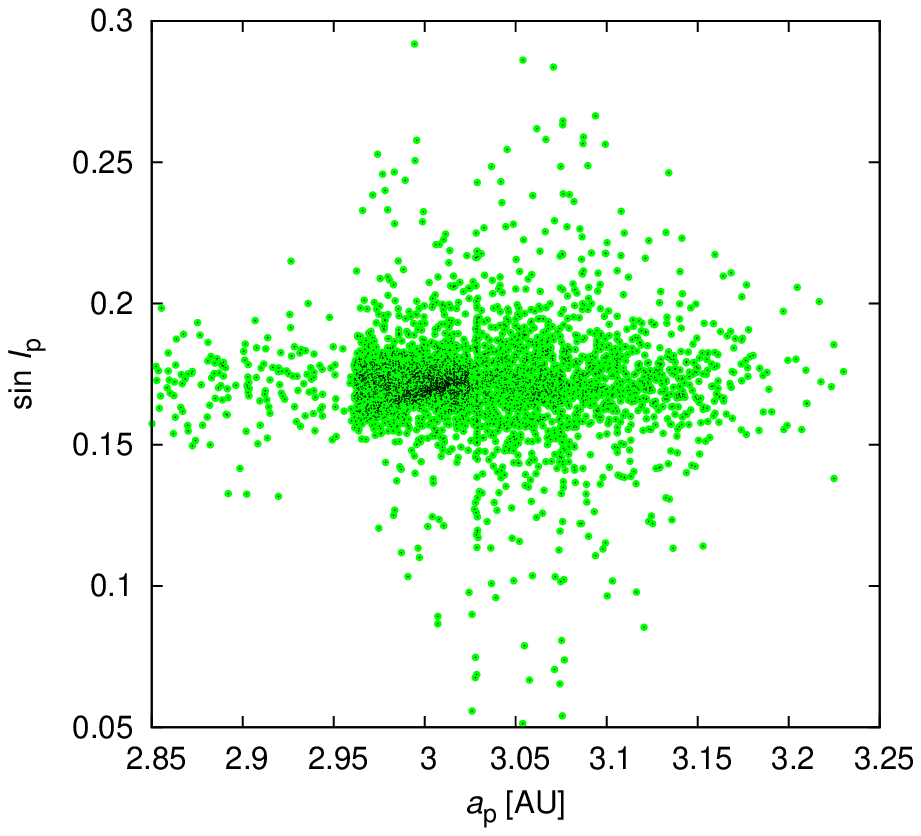}
\caption{Top panel:
the proper semimajor axis~$a_{\rm p}$ vs proper inclination~$\sin I_{\rm p}$
for all asteroids in the broad surroundings of Eos family.
The range of proper eccentricities is $e_{\rm p} \in (0.0; 0.3)$.
If they have colour data in the SDSS MOC4 catalogue \citep{Parker_etal_2008Icar..198..138P},
the colours correspond to indices $a^\star$, $i-z$
which are closely related to taxonomy,
namely blue is close to C-complex taxonomy,
red to S-complex,
and magenta to K-type.
The whole sample contains 18\,471 asteroids.
There are other prominent families visible:
Hygeia (C-type, bottom-right),
Veritas (C, next to Eos),
Tirela (S, upper right),
Telramund (S, below Eos);
a close inspection would show 32 families in total!
Bottom panel:
the same plot for a typical outcome of N-body simulations,
assuming a disruption of a parent body,
ejection of fragments with some velocity field,
and their long-term dynamical evolution due to
gravitational perturbations,
resonances,
chaotic diffusion,
the Yarkovsky effect,
the YORP effect, etc.
The two panels are not directly comparable.}
\label{Ktypes_ALL_ai_sloan}
\end{figure}

A rigorous comparison of observations {\em versus\/} simulations of asteroid families
is a~rather difficult task, especially when the observations look like Figure~\ref{Ktypes_ALL_ai_sloan}.
Observed proper elements $a_{\rm p}$, $e_{\rm p}$, $\sin I_{\rm p}$,
supplied by physical data (colour indices $a^\star$, $i-z$ in this case),
show a~complicated structure of the Eos family, halo,
together with many neighbouring families, overlapping halos, and background asteroids, of course.
The hierarchical clustering method alone (HCM, \citealt{Zappala_etal_1995Icar..116..291Z})
is then practically useless.

Family identification itself affects dynamical studies and {\em vice versa\/}.
We would need the family to determine initial conditions.
On the other hand, we would need a dynamical study to understand
whereever family members could be.
There are several well-known weaknesses of HCM, which were demonstrated
e.g. in a `crime-scene' Fig.~8 of \cite{Nesvorny_etal_2015aste.book..297N}.
The HCM needs a free parameter,
either the cutoff velocity $v_{\rm cut}$,
or the quasi-random level ${\rm QRL}$.
It is also unable to associate halos.
Last but not least, the background is never precisely uniform
what can be clearly seen at the edges of currently stable zones,
close or inside gravitational resonances,
or even in stable zones where the population was deteriorated
by dynamical processes in the distant past (cf.~Cybele region;
\citealt{Carruba_etal_2015MNRAS.451..244C}).

On the other hand, synthetic families evolve in the course
of simulation and loose their members, consequently we should use
a~variable $v_{\rm cut}$, but its optimal value is again
generally unknown. No direct comparison is thus possible.

That is a motivation for our work. We describe a method
suitable to study 3-dimensional shapes of asteroid families,
taking into account all proper orbital elements,
including possibly non-uniform background,
and matching the size-frequency distribution at the same time.
Our method still relies on a preliminary selection
of observed asteroids according to their colours (or albedos)
to suppress -- but {\em not\/} fully exclude -- interlopers.
A comparison of the observed asteroids with an output of N-body simulation
is performed by means of counting the bodies in proper-element 'boxes',
and a suitable $\chi^2$ metric. Because we are forced to select
synthetic asteroids randomly (a Monte-Carlo approach),
we can expect some stochasticity of the results.

We present an application to the Eos family
(family identification number, FIN = 606),
one of the most studied families to date,
mentioned already by \cite{Hirayama_1918AJ.....31..185H}.
Together with our previous works
\citep{Vokrouhlicky_etal_2006Icar..182...92V,Broz_Morbidelli_2013Icar..223..844B},
this paper forms a long-term series focused on its long-term evolution.
We use up-to-date catalogues of proper elements \citep{Knezevic_Milani_2003A&A...403.1165K},
and brand new spin data \citep{Hanus_etal_2018Icar..299...84H}.

Let us recall that the Eos family is of~K taxonomic type,
while the background is mostly C type.
\cite{MotheDiniz_etal_2008Icar..195..277M} suggested either
a partially differentiated parent body,
with meteorite analogues CV, CO or R;
or a~undifferentiated one, with CK analogues.
There was a discovery of a recent breakup of (6733) 1992~EF
\citep{Novakovic_Tsirvoulis_2014acm..conf..388N},
belonging to the family core,
what makes Eos even more interesting for space weathering studies,
because we may see both old (1.3\,Gyr) and young (4\,Myr) surfaces.

\section{Methods}\label{sec:methods}

Before we proceed with the description of the method,
let us explain three problems we have to solve
and describe the underlying dynamical model.

\subsection{Problem 1: Selection of asteroids}\label{sec:problem1}

In principle, we can select any subset of asteroids
(e.g. by using SDSS colour data, or WISE albedo data) to decrease a contamination
by interlopers, or an overlap with other families in the neighbourhood
\citep{Parker_etal_2008Icar..198..138P,Masiero_etal_2011ApJ...741...68M};
an approach also used in a multidomain HCM \citep{Carruba_etal_2013MNRAS.433.2075C}.
We can also simulate any subset at will, but we should definitely check surroundings
where the bodies can be scattered to, because this may be a key constraint.

\begin{figure}
\centering
\includegraphics[height=6cm]{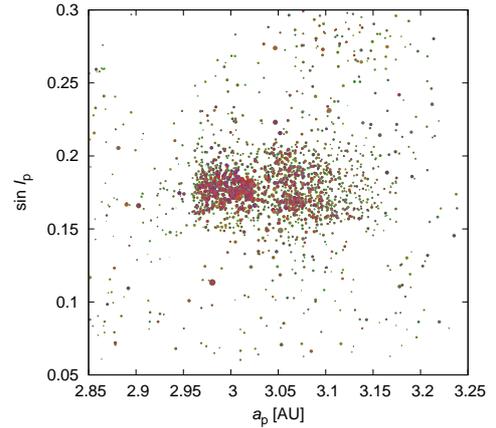}
\caption{K-type asteroids selected from Figure~\ref{Ktypes_ALL_ai_sloan},
with known colour indices
$a^*\! \in (0.0; 0.1)$,
$i-z \in (-0.03; 0.08)$.
The visual geometric albedo had to be $p_V > 0.07$ (or unknown).
This subset is much more homogeneous and contains 1\,991 asteroids.
No other prominent families except Eos can be seen;
the only exception may be some contamination by Tirela (upper right)
due to inherent photometric noise.
This subset seems already suitable for a~comparison with N-body simulations.}
\label{Ktypes_ai_sloan}
\end{figure}

For Eos family, it is easy because of its distinct K taxonomic type
which is defined for our purposes in terms of the SDSS colour indices
$a^*\! \in (0.0, 0.1)$,
$i-z \in (-0.03, 0.08)$,
and the geometric albedo $p_V > 0.07$ (if known in WISE or IRAS catalogues).
If only colours are known, we select the asteroids according to them,
and assume their $p_V = 0.158$ which corresponds to the median value of Eos members.
As a result, only 1/10$^{\rm th}$ of asteroids remain, but this is still sufficient
(Figure~\ref{Ktypes_ai_sloan}).
Practically all other families have disappeared,
the background is much more uniform.
The only exception may be some contamination from the Tirela family
(seen as a concentration in the upper right corner of Fig.~\ref{Ktypes_ai_sloan}),
arising from a photometric noise on S-type asteroids,
and a gap at large $\sin I_{\rm p} > 0.25$.

\begin{figure}
\centering
\includegraphics[height=5.5cm]{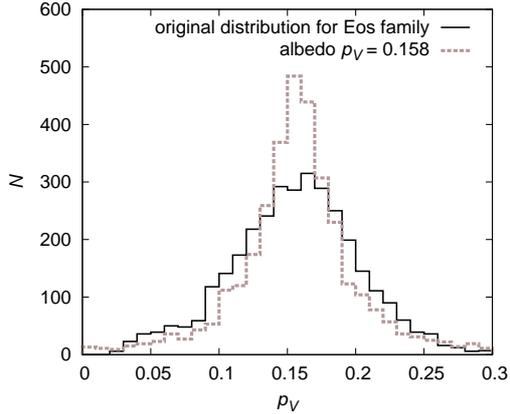}
\caption{The observed differential distribution of visual geometric albedos $p_V$
for the Eos family from the WISE catalogue \citep{Masiero_etal_2011ApJ...741...68M} (black solid),
and for the same set of bodies with $p_V$ values assigned randomly,
assuming a Gaussian distribution
with a constant mean $\bar p_V = 0.158$, and
1-$\sigma$ uncertainty declared in the catalogue (dashed gray).
The widths of the two distributions are similar,
so using the constant $\bar p_V$ (if unknown) is {\em not\/} a~poor approximation.}
\label{albedo}
\end{figure}

Regarding the homogeneity of albedos, the WISE data exhibit a wide distribution,
and we should check whether it can be related to a heterogeneous parent body.
The uncertainties~$\sigma_{\!p}$ arise mainly from photon noise,
and NEATM model systematics. In a statistical sense, even the single albedo value
$\bar p_V = 0.158$ would result in a relatively wide distribution
because $\sigma_{\!p}$ values are relatively large,
which is demonstrated in Figure~\ref{albedo},
where we used the $\sigma$'s of {\em individual\/} measurements
together with the (constant) $\bar p_V$
to randomly generate the new distribution of $p_V$'s.
The Eos family thus seems homogeneous rather than heterogeneous.

\subsection{Problem 2: Size-frequency distribution}\label{sec:problem2}

\begin{figure}
\centering
\includegraphics[height=5cm]{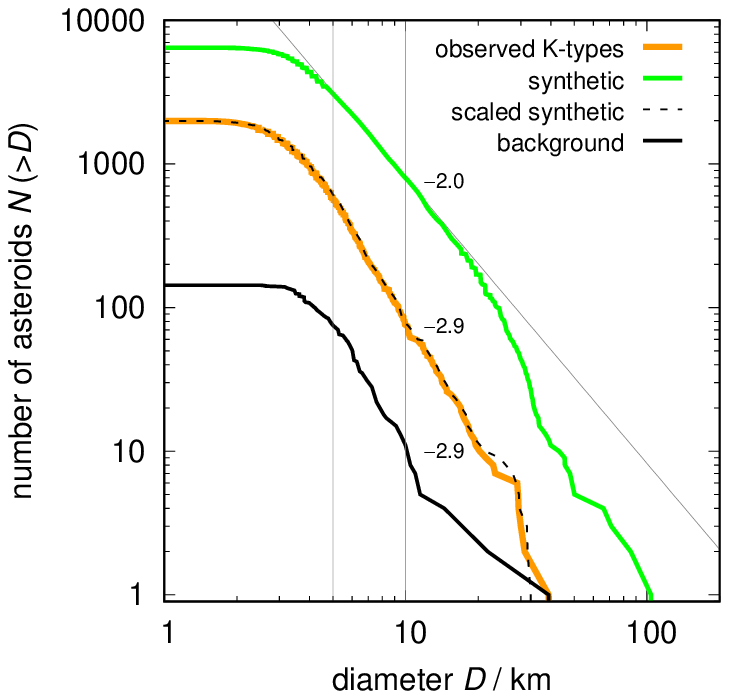}
\includegraphics[height=5cm]{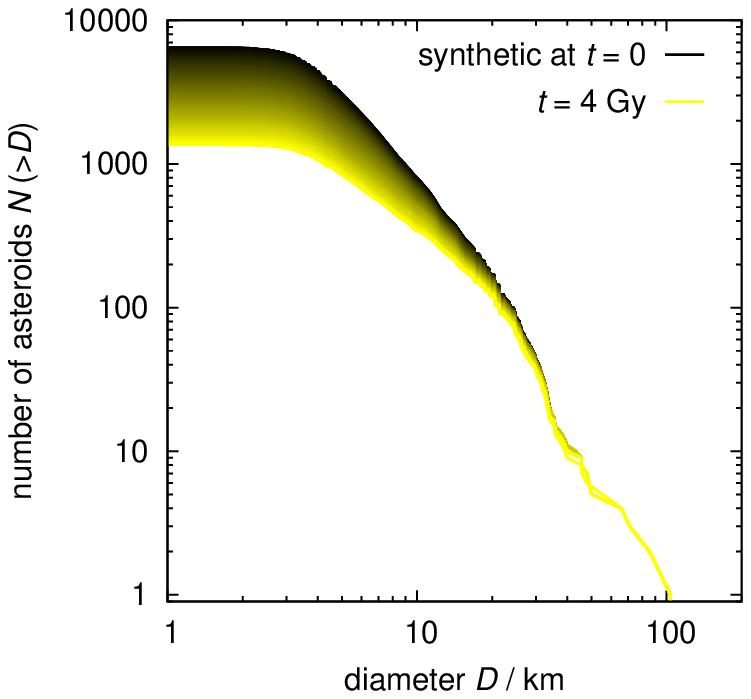}
\caption{Top panel:
the cumulative size-frequency distribution (SFD) of the observed K-type asteroids (orange),
the synthetic SFD at the beginning of N-body simulation (green),
the {\em scaled\/} synthetic SFD constructed by a random selection of bodies
so that it matches the observed one (dotted black; hard to distinguish from orange),
and the background SFD (black).
Bottom panel:
an evolution of the synthetic SFD in the course of an N-body simulation,
from time $t = 0$ up to 4\,Gyr,
which is indicated by changing colours (black$\,\rightarrow\,$yellow).
These changes (due to a dynamical decay) require scaling at every time step.}
\label{size_distribution_BACKGROUND2}
\end{figure}

The size-frequency distributions (SFDs) should match
for both the observed and synthetic populations,
but the latter changes in the course of time (Figure~\ref{size_distribution_BACKGROUND2}).
In order to compare apples with apples, we have to scale the SFD.
In other words, we randomly select the same number of synthetic bodies
(together with their orbits, of course) as the number of observed bodies,
in each of prescribed size bins $(D, D+{\rm d}D)$.
Let us emphasize we do not rely on the assumption of a constant SFD.%
\footnote{In principle, we can estimate the original SFD of the family
but it is not our goal here. The overall change of slope due to
dynamical decay (for selected~$t$) can be estimated already from
Fig.~\ref{size_distribution_BACKGROUND2}.}
To this point, it is definitively useful to start with
a larger number of synthetic bodies, so that
we still have more than observed at the end of simulation.

This random selection of synthetic asteroids to match the SFD of observed asteroids
is needed at every single output time step of the simulation.
Even multiple selections at one time step might be useful.
This way, we would naturally account for an additional
(and often neglected) uncertainty which arises from the fact
we always choose the initial conditions from some underlying distributions
(e.g. from a prescribed velocity field), but we cannot be
absolutely sure that our single selection is not a lucky fluke.

\subsection{Problem 3: Non-uniform background}\label{sec:problem3}

A background has to be accounted for otherwise it is essentially impossible
to explain a lot of bodies far from the family.
First, we need to find some observed background,
not very far from the family;
in our case, a suitable population seems to be at
$\sin I_{\rm p} \in (0.06; 0.12)$ and $(0.24; 0.30)$.
It has its own size-frequency distribution,
and we should use the same SFD for the synthetic background.
As a first approximation, we model the background
as a random uniform distribution in the space of proper elements.

\begin{figure}
\centering
\includegraphics[height=6cm]{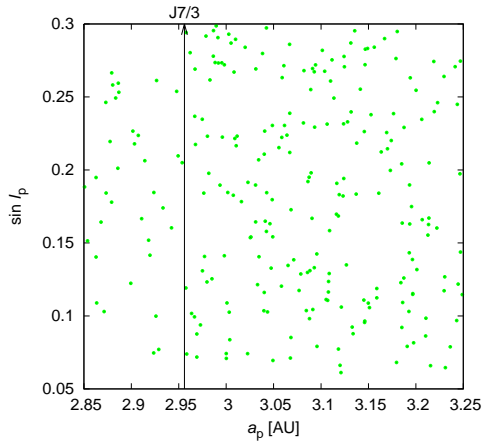}
\caption{A synthetic background generated as a random uniform distribution
in proper orbital elements $a_{\rm p}, e_{\rm p}, \sin I_{\rm p}$,
with the same size-distribution as the observed background.
In this example, the number densities below and above
the 7/3 mean-motion resonance with Jupiter at 2.956\,au
are different (by a factor of 2), because this resonance
separates two distinct zones of the main belt.}
\label{ai_background}
\end{figure}

However, Murphy's law for backgrounds states: {\em The background is never uniform.\/}
Especially below and above the 7/3 mean-motion resonance with Jupiter
we can expect a difference (see the example in Figure~\ref{ai_background}).

Again, there is a non-negligible stochasticity.
We shall at least try a~different random seed.
The number density of background objects can be also treated
as a free parameter. There is also {\em a priori\/} unknown systematic contamination
by neighbouring families, but this is not necessarily present
right `under' the Eos family.

\subsection{Dynamical model}\label{sec:dynamics}

Our dynamical model was described in detail in \cite{Broz_etal_2011MNRAS.414.2716B}.
We briefly recall it contains
a modified SWIFT integrator \citep{Levison_Duncan_1994Icar..108...18L, Laskar_Robutel_2001CeMDA..80...39L},
both the diurnal and seasonal Yarkovsky thermal effects \citep{Vokrouhlicky_1998A&A...335.1093V, Vokrouhlicky_Farinella_1999AJ....118.3049V},
which induce a semimajor axis drift ${\rm d}a/{\rm d}t$;
all mean-motion and secular resonances,
captures and corresponding drifts ${\rm d}e/{\rm d}t$, ${\rm d}I/{\rm d}t$,
the YORP effect,
changing the spin rate $\omega$ and the obliquity~$\gamma$ \citep{Capek_Vokrouhlicky_2004Icar..172..526C},
with the efficiency parameter $c_{\rm YORP} = 0.33$ \citep{Hanus_etal_2011A&A...530A.134H},
simplified collisional reorientations by means of a prescribed
time scale dependent on size~$D$ \citep{Farinella_etal_1998Icar..132..378F},
random period changes due to mass shedding after reaching the critical spin rate~$\omega_{\rm crit}$ \citep{Pravec_Harris_2000Icar..148...12P},
and suitable digital filters for computations of mean and proper elements
\citep{Quinn_etal_1991AJ....101.2287Q, Sidlichovsky_Nesvorny_1996CeMDA..65..137S}.

\begin{figure}
\centering
\includegraphics[width=8.0cm]{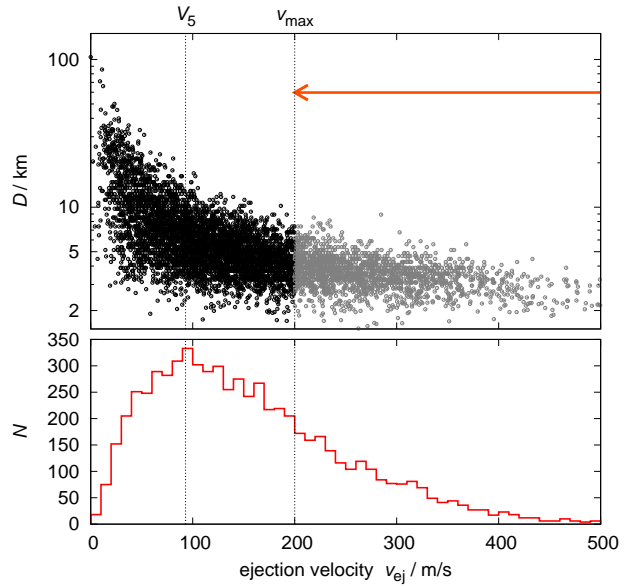}
\caption{Top panel: the dependence of the ejection velocity~$v_{\rm ej}$
on the diameter~$D$ for our synthetic bodies.
The value $V_5 = 93\,{\rm m}\,{\rm s}^{-1}$ denotes the dispersion
of velocity components for $D_5 = 5\,{\rm km}$ bodies.
In specific cases (Sec.~\ref{badfit1}), we select only bodies
with velocities smaller than some maximum value, $v_{\rm ej} < v_{\rm max}$.
Bottom panel: the corresponding histogram of~$v_{\rm ej}$.}
\label{genveld2}
\end{figure}

Initial conditions are kept as simple as possible.
We assume an isotropic disruption,
with the ejection velocity components Gaussian,
with the dispersion proportional to $1/D$, and
$V_5 = 93\,{\rm m}\,{\rm s}^{-1}$
for $D_5 = 5\,{\rm km}$,
an estimate based on our previous work \citep{Vokrouhlicky_etal_2006Icar..182...92V}.
Consequently, the distrubution of the velocity magnitude $|\vec v_{\rm ej}|$ is Maxwellian (see Figure~\ref{genveld2}).
We start with 6\,545 synthetic bodies,
with the SFD covering $D \in (1.5; 100)\,{\rm km}$.
Spins are also isotropic and periods uniform, $P \in (2; 10)\,{\rm h}$.

The thermal parameters remain the same as in our previous works:
the bulk density $\rho = 2\,500\,{\rm kg}\,{\rm m}^{-3}$,
the surface density $\rho = 1\,500\,{\rm kg}\,{\rm m}^{-3}$,
the conductivity $K = 0.001\,{\rm W}\,{\rm m}^{-1}\,{\rm K}^{-1}$,
the specific capacity $C = 680\,{\rm J}\,{\rm kg}^{-1}$,
the Bond albedo $A = 0.1$,
the infrared emissivity $\epsilon = 0.9$.
For simplicity, we assumed these parameters to be constants,
although some of them may be size-dependent (as $K$ in \citealt{Delbo_etal_2015aste.book..107D}),
or temperature-dependent \citep{Anderson_etal_1991JGR....9618037A}.

The free parameters of our model are
the maximum of velocity distribution $v_{\rm max}$ (Fig.~\ref{genveld2}),
the true anomaly~$f_{\rm imp}$, and
the argument of pericentre $\omega_{\rm imp}$ at the time of impact,
which are interrelated by means of the Gauss equations.
We may be forced to tune also other osculating orbital elements of the parent body,
but for the moment we take those of (221) Eos as the nominal case.

Among the fixed parameters is the bulk density~$\rho$.
Usually, the age scales linearly with $\rho$
due to the non-gravitational accelerations.
Theoretically, if there are both gravitational and non-gravitational
accelerations acting at the same time
(e.g. Yarkovsky drift in~$a$ and chaotic diffusion in~$e$)
we may be able to break this degeneracy.
However, based on our previous experience,
we do not expect this for Eos.
Neighbouring Veritas may be more suitable for this approach, by the way.
Alternatively, one can use collisional models
which exhibit a different scaling with $\rho$
(cf.~Sec.~\ref{sec:collisional}).

We integrate the equations of motion with
the time step $\Delta t = 91\,{\rm d}$, and
the time span $4\,{\rm Gyr}$.
The output time step after computations of mean elements,
proper elements, and final running-window filter
is $\Delta t_{\rm out} = 10\,{\rm Myr}$.


\subsection{Black-box method}

We can eventually proceed with a so-called `black-box' method (see Figure~\ref{eos-5_AE_RESCALESFD_BACKGROUND_ae_syn_0000})%
\footnote{see \url{http://sirrah.troja.mff.cuni.cz/~mira/eos/eos.html} for an implementation in Python}%
: (i)~we choose 180~boxes with $\Delta a = 0.0243\,{\rm au}$, $\Delta e = 0.025$, $\Delta\sin I = 0.240$
in our case aligned with the J7/3 and J9/4 resonances%
\footnote{possibly also in~$D$}%
;
(ii)~count the numbers of observed asteroids located in these boxes;
(iii)~compute the observed incremental SFD globally, in the full domain;
(iv)~compute the background incremental SFD globally;
(v)~at every single output time step 
we compute the synthetic incremental SFD globally again
(saving also lists of bodies in the respective size bins);
(vi)~for every single size bin $(D, D+{\rm d}D)$
we draw a~synthetic background population of $N_{\rm bg}$ bodies from a random uniform distribution
(in the whole range of $a_{\rm p}, e_{\rm p}, \sin I_{\rm p}$);
if the volume where the background was selected
differs from our volume of interest, we have to use a suitable factor, i.e. $fN_{\rm bg}$;
(vii)~we {\em scale\/} the synthetic SFD to the observed one
by randomly choosing $N_{\rm obs}-fN_{\rm bg}$ bodies from the lists above;
(viii)~we count the numbers of all synthetic asteroids located in the boxes;
(ix)~finally, we compute the metric
\begin{equation}
\chi^2 = \sum_{i=1}^{N_{\rm box}} {(N_{{\rm syn}\,i}-N_{{\rm obs}\,i})^2\over \sigma_{{\rm syn}\,i}^2 + \sigma_{{\rm obs}\,i}^2}\,,
\end{equation}
where the uncertainties are assumed Poisson-like, $\sigma = \sqrt{N}$.
Using both $\sigma_{\rm obs}$ and $\sigma_{\rm syn}$ in the
denominator prevents `extreme' $\chi^2$ contributions
in boxes where $N_{\rm obs} \to 0$.
We shall keep in mind though the corresponding
probability distribution of $\chi^2$ may be somewhat skewed.
There is some freedom related to the box sizes (binning),
but within the limits of meaningfulness (neither a single box
nor zillions of boxes), the method should give statistically
comparable results as we always analyse the same information.

Unlike traditional simplified methods fitting an envelope to $(a_{\rm p}, H)$ or $(a_{\rm p}, 1/D)$,
we shall obtain not only an upper limit for the age,
but also a lower limit.


\section{Results}\label{sec:results}

Hereinafter, we discuss not only the best-fit model,
but also several bad fits which are actually more important,
because the `badness-of-fit' assures a solid conclusion
about the Eos family.

\subsection{The nominal model}

\begin{figure*}
\centering
\setlength{\tabcolsep}{2pt}
\begin{tabular}{cccc}
$t = 0\,{\rm Myr}$ & $t = 1340\,{\rm Myr}$ & $t = 3980\,{\rm Myr}$ \\[3pt]
\includegraphics[height=3.5cm]{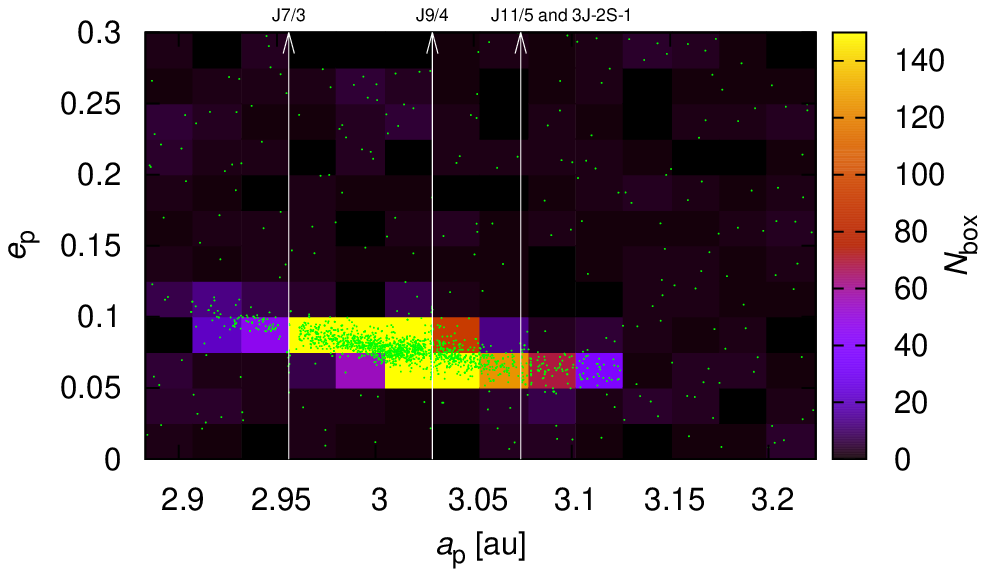} &
\includegraphics[height=3.5cm]{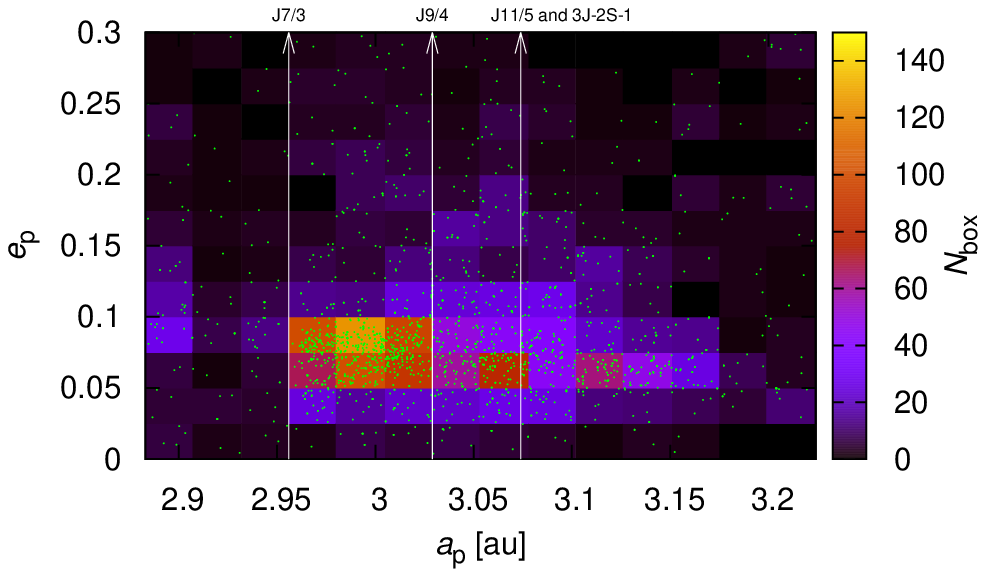} &
\includegraphics[height=3.5cm]{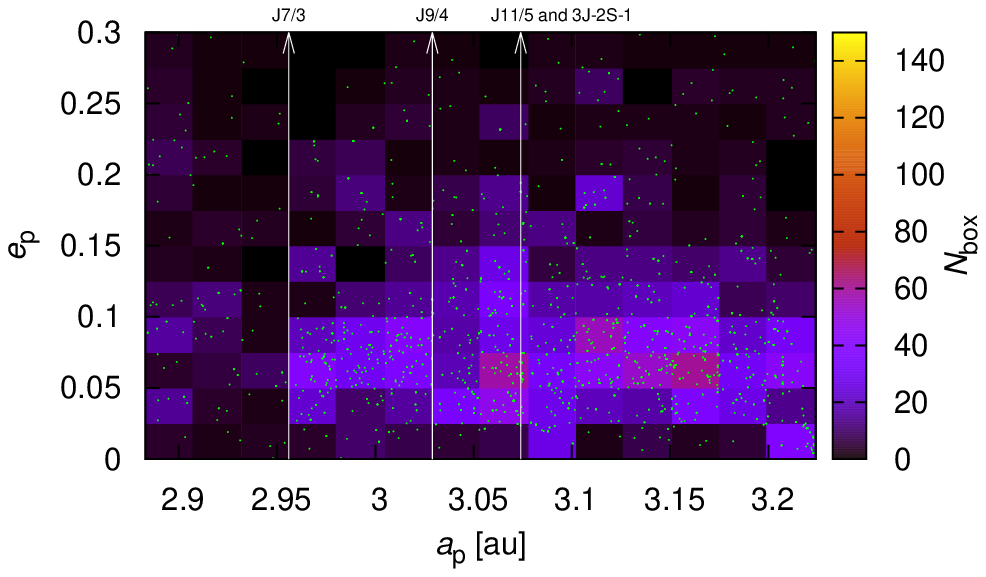} \\
&
\includegraphics[height=3.5cm]{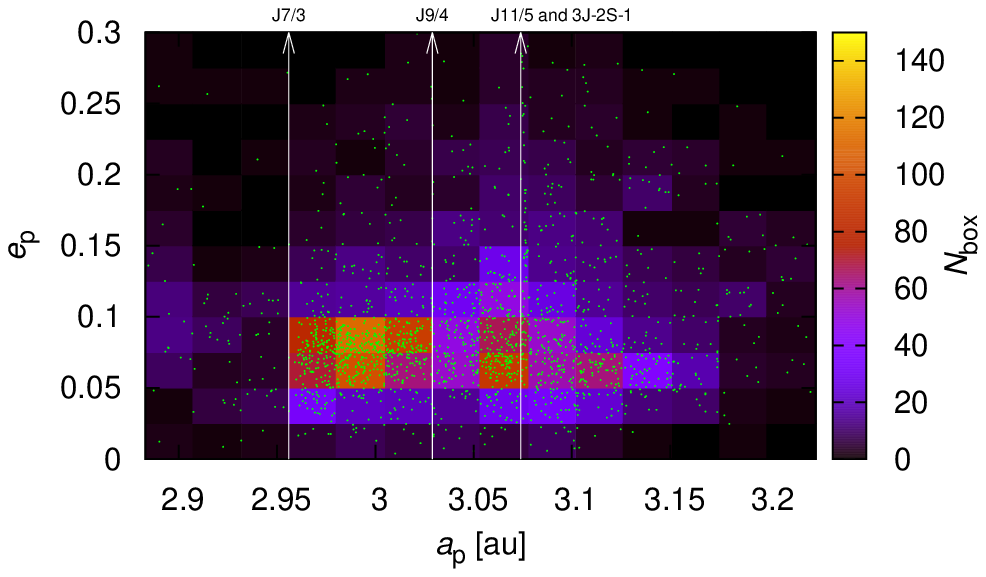} &
\raise1.8pt\hbox{\includegraphics[height=3.35cm]{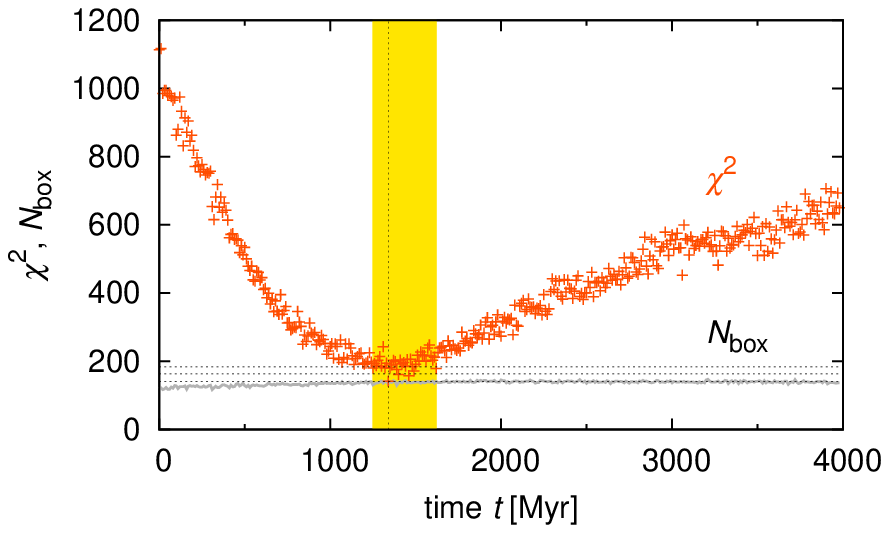}} \\
& observed & $\chi^2$ metric \\
\end{tabular}
\caption{
The proper semimajor axis~$a_{\rm p}$ vs proper eccentricity~$e_{\rm p}$
for the nominal simulation scaled to the observed SFD (as described in the main text)
(top row).
Bodies are plotted as green dots.
Colours correspond to the number of bodies in 180~boxes, outlined by 
$\Delta a = 0.0234\,{\rm au}$,
$\Delta e = 0.025$.
The range of inclinations is always $\sin I_{\rm p} \in (0.06; 0.30)$.
Positions of major mean-motion and 3-body resonances are also indicated
(J7/3, J9/4, J11/5, and $3{\rm J}-2{\rm S}-1$).
The $z_1$ secular resonance goes approximately from the lower-left corner to the upper-right.
There are the initial conditions (left column),
the best-fit at $t = 1340\,{\rm Myr}$ (middle),
the end of simulation (right);
as well as the observations (bottom middle), and
the respective $\chi^2$ metric compared to the actual number of boxes $N_{\rm box}$ (bottom right).
The correspondence between the best-fit and the observations is surprisingly good,
with
$\chi^2 = 141$,
$N_{\rm box} = 134$ (not all boxes are populated),
and $\chi^2 \simeq N_{\rm box}$.
The 1-$\sigma$, 2-$\sigma$ and 3-$\sigma$ levels (dotted lines)
and the inferred 3-$\sigma$ uncertainty of the age (yellow strip) are indicated too.}
\label{eos-5_AE_RESCALESFD_BACKGROUND_ae_syn_0000}
\end{figure*}

The nominal model is presented in Figure~\ref{eos-5_AE_RESCALESFD_BACKGROUND_ae_syn_0000}.
We focus on the proper semimajor axis~$a_{\rm p}$ vs proper eccentricity~$e_{\rm p}$ distribution,
having only one box in inclination $\sin I_{\rm p}$.
The initial conditions (top left) are so different from the observations (bottom middle)
it is almost hopeless to expect a good fit anytime in the future.
However, at around $t = 1.3\,{\rm Gyr}$ the situation suddenly changes (top middle);
it is almost unbelievable that the synthetic family is so similar to the observations!
The final state (top right) is again totally different.
The $\chi^2$ reaches values as low as $N_{\rm box}$,
so we may consider the best fit to be indeed reasonable.
The age interval is $t = (1.3\pm0.3)\,{\rm Gyr}$.
Let us emphasize that the fit so good only because
we carefully accounted for all three problems outlined in Section~\ref{sec:methods}.

\subsection{Bad fit 1: Ejection velocity tail}\label{badfit1}

Because our sample is 3~times larger than the observed sample,
we can easily resample our synthetic bodies without actually
computing the N-body simulation anew, e.g. selecting only
those with low ejection velocity $v_{\rm ej} < 200\,{\rm m}\,{\rm s}^{-1}$.
Consequently, all bodies are initially located above the J7/3 resonance,
and below the J11/5.

\begin{figure}
\centering
\begin{tabular}{c}
$t = 1430\,{\rm Myr}$ \\
\includegraphics[width=7.0cm]{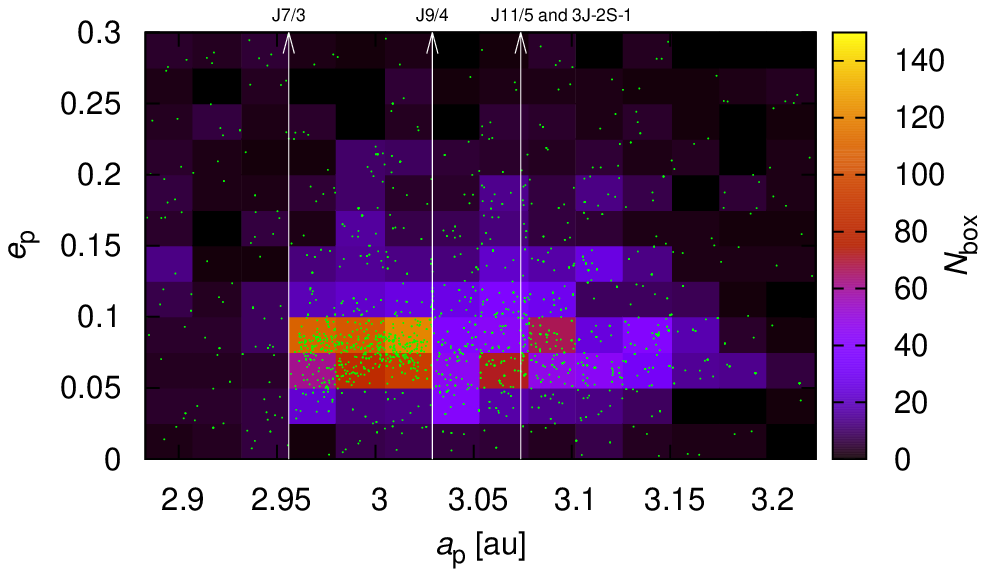} \\
\kern-1.0cm\hbox{\includegraphics[width=6.1cm]{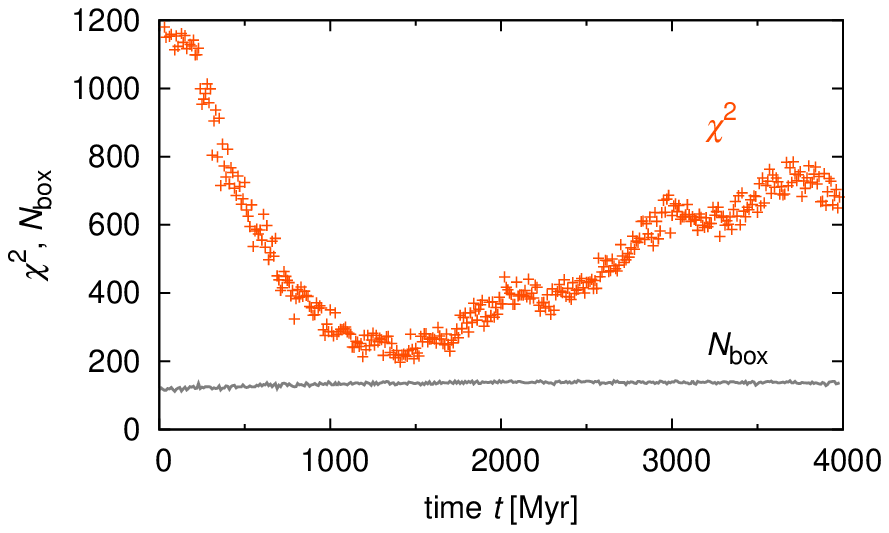}} \\
\end{tabular}
\caption{Bad fit 1:
the proper semimajor axis~$a_{\rm p}$ vs proper eccentricity~$e_{\rm p}$ (top panel),
and the temporal evolution of $\chi^2$ (bottom panel)
for a subset of bodies with the ejection velocities $v_{\rm ej} < 200\,{\rm m}\,{\rm s}^{-1}$,
i.e. without the tail of the distribution.
Initially, all bodies were located above the J7/3 resonance.
Observations were shown in Fig.~\ref{eos-5_AE_RESCALESFD_BACKGROUND_ae_syn_0000} (bottom middle).
The `best-fit' at $t = 1430\,{\rm Myr}$,
with $\chi^2 = 197$,
$N_{\rm box} = 134$,
is much worse than the nominal case.
The number of bodies below the J7/3 resonance is too low.
Consequently, the velocity tail is needed to get a better fit.}
\label{eos-5_VLT200_AE_RESCALESFD_BACKGROUND_ae_syn_1410}
\end{figure}

Using the same post-processing as above we arrived at Figure~\ref{eos-5_VLT200_AE_RESCALESFD_BACKGROUND_ae_syn_1410}.
It is clear that the `best fit' is actually a bad fit
compared to the nominal model.
The notable differences are below the J7/3 resonance,
and above the J11/5 where the numbers of bodies
are never sufficient to match the observations
(cf.~Fig.~\ref{eos-5_AE_RESCALESFD_BACKGROUND_ae_syn_0000}, bottom middle).

It is worth to note there is a~small family just below the J7/3
resonance, namely (36256) 1999 XT$_{17}$ (FIN 629).
\cite{Tsirvoulis_etal_2018Icar..304...14T} discovered a~link to Eos
by analysing the overall V-shape in the semimajor axis~$a_{\rm p}$
vs the absolute magnitude~$H$ diagram.
It seems aligned with the original velocity field of the Eos family
--- it has the same $\sin I_{\rm p}$ as the family core,
but slightly larger $e_{\rm p} \simeq 0.1$, because of the `ellipse'
in $(a_{\rm p}, e_{\rm p})$ visible in Fig.~\ref{eos-5_AE_RESCALESFD_BACKGROUND_ae_syn_0000} (top left).
We thus conclude, (36256) family is actually a {\em remnant\/}
of the original velocity field.

If this is true, it may further contribute to the contamination
of the `pristine zone' between the J7/3 and J5/2 resonances,
apart from low-probability crossings of the former resonance.
This region was analysed by \cite{Tsirvoulis_etal_2018Icar..304...14T},
where authors carefully subtracted the contribution of all families
(including Eos), extracted the SFD of remaining background asteroids
and computed the slope of the primordial (post-accretion) SFD.

\subsection{Bad fit 2: Parent body inclination}

\begin{figure}
\centering
\begin{tabular}{c}
$t = 1270\,{\rm Myr}$ \\
\includegraphics[width=7.0cm]{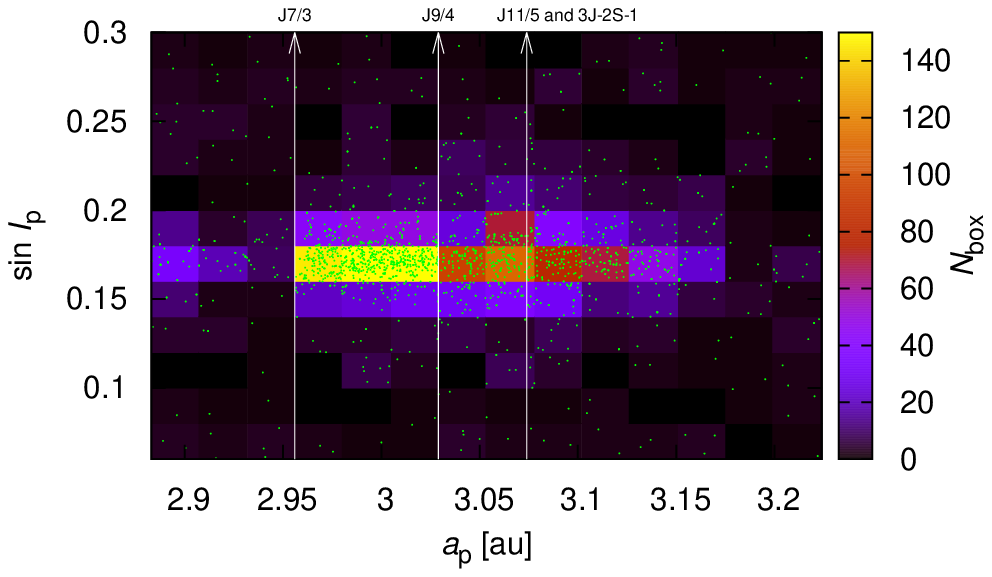} \\
\kern-1.0cm\hbox{\includegraphics[width=6.1cm]{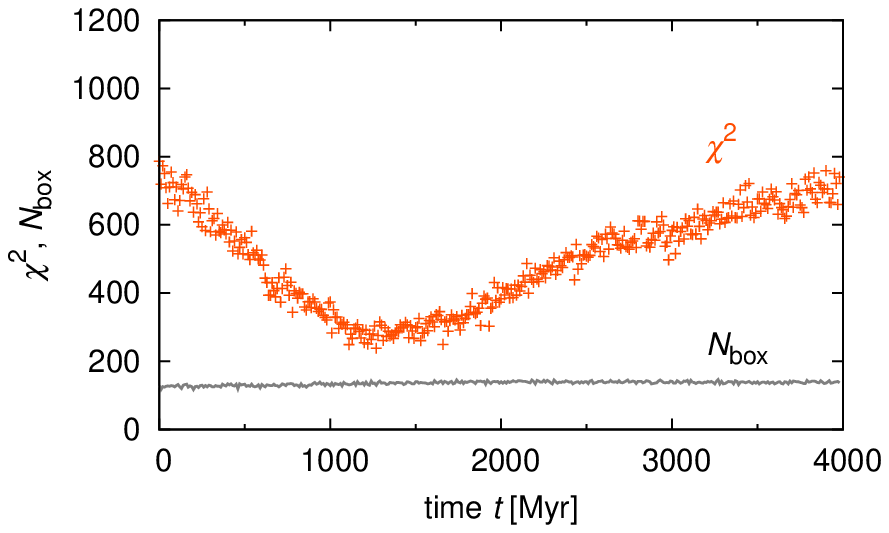}} \\
\end{tabular}
\caption{Bad fit 2:
the proper semimajor axis~$a_{\rm p}$ vs proper inclination~$\sin I_{\rm p}$
for the synthetic population (top panel),
and the temporal evolution of $\chi^2$ (bottom panel).
The boxes are consequently different,
$\Delta a = 0.0243\,{\rm au}$,
$\Delta\sin I = 0.02$,
$e_{\rm p} \in (0.0; 0.3)$,
so is the resulting `best-fit' value
$\chi^2 = 238$,
$N_{\rm box} = 130$.
The parent body would have to be shifted in inclination
by approximately $0.005\,{\rm rad}$ to get a better fit.}
\label{eos-5_AI_BACKGROUND_PLUS0_000_FACTOR2_ai_syn_1270}
\end{figure}

If we look on contrary on the proper semimajor axis~$a_{\rm p}$
vs proper inclination~$\sin I_{\rm p}$ distribution
(Figure~\ref{eos-5_AI_BACKGROUND_PLUS0_000_FACTOR2_ai_syn_1270})
there is a problem with the nominal model.
Inclinations are all the time too low
(and the $\chi^2$ too high compared to $N_{\rm box}$).
This would affect a 3-dimensional fit too, of course.

Nevertheless, it seems sufficient to adjust the inclination
by approximately $0.005\,{\rm rad}$ to get a significantly better fit,
$\chi^2$ decreased from 238 down to 181.
This seems still too high wrt.~130, but this approach is possibly too simplified,
because we only shifted the output data.
In reality, the resonances (in particular the $z_1$) do not shift at all,
they are determined by the positions of giant planets,
and we should perform the N-body integration anew to obtain
a correct $(a_{\rm p}, \sin I_{\rm p})$ distribution.

\subsection{Bad fit 3: True anomaly $f_{\rm imp} < 120^\circ$}

\begin{figure}
\centering
\begin{tabular}{c}
$t = 1280\,{\rm Myr}$ \\
\includegraphics[width=7.0cm]{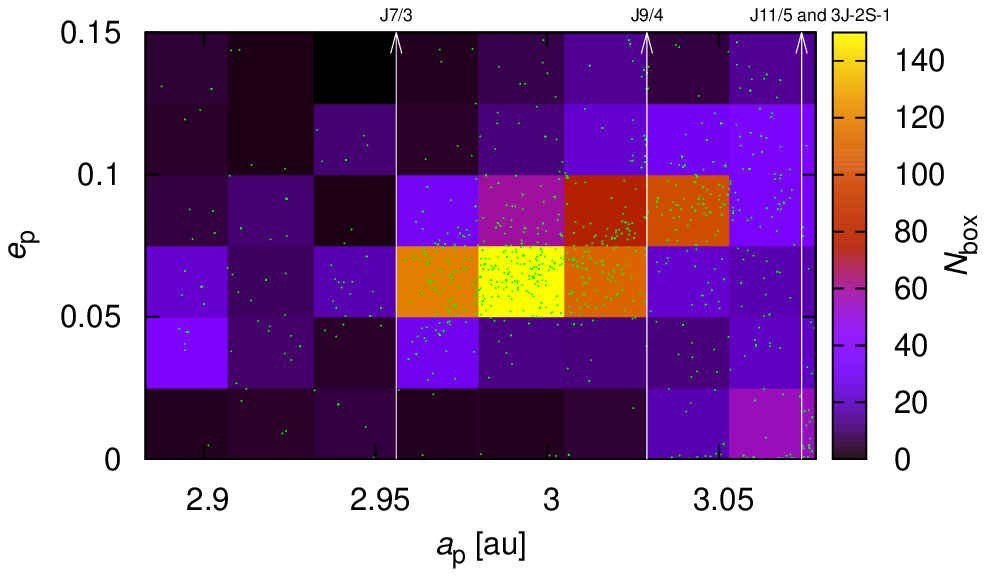} \\
\kern-1.0cm\hbox{\includegraphics[width=6.1cm]{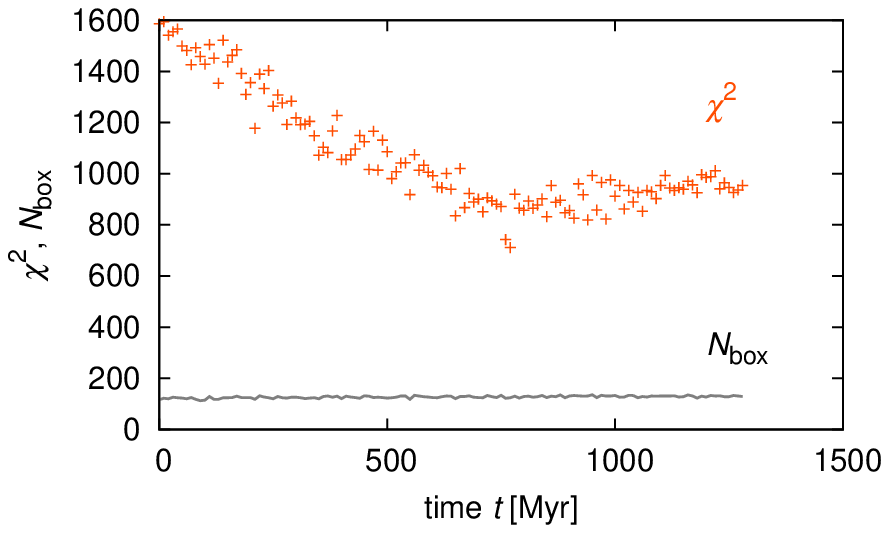}} \\
\end{tabular}
\caption{Bad fit 3:
a detail of the proper semimajor axis~$a_{\rm p}$ vs proper eccentricity~$e_{\rm p}$ (top panel),
and the temporal evolution of $\chi^2$ (bottom panel)
for the simulation with the true anomaly at the time of impact $f_{\rm imp} = 0^\circ$,
and the argument of perihelion $\omega_{\rm imp} = 30^\circ$.
The `best-fit' $\chi^2 = 711$ is so high compared to $N_{\rm box} = 124$
that the simulation was not computed up to $4000\,{\rm Myr}$.
The value has to be $f \gtrsim 120^\circ$ to get a better fit.}
\label{eos-4_AE_RESCALESFD_BACKGROUND_ae_syn_1280}
\end{figure}

To demonstrate the sensitivity of our `black-box' method
with respect to the impact parameters, we present an alternative N-body simulation
which started with the true anomaly $f_{\rm imp} = 0^\circ$.
The orientation of the ellipse is then the opposite
and there is practically no chance for a good fit
(see Figure~\ref{eos-4_AE_RESCALESFD_BACKGROUND_ae_syn_1280}).

All the time, there is a serious mismatch within the family core,
it is impossible explain the observed bodies in the boxes with
$a_{\rm p} \simeq 2.97\,{\rm au}$, and
$e_{\rm p} \simeq 0.08$.
Generally, it is surprising that even $1.3\,{\rm Gyr}$ after the impact,
there are clear traces of the original velocity field!
As already reported in \cite{Broz_Morbidelli_2013Icar..223..844B},
the `true' true anomaly should be $f > 120^\circ$.
Another example of such traces (in inclination)
is Koronis family \citep{Carruba_etal_2016Icar..271...57C}.

\section{Conclusions}\label{sec:conclusions}

Let us conclude, it is important
to use a suitable selection of asteroids,
match the size-frequency distributions,
and account for the background population,
when comparing N-body simulations with observations.
To this point, we presented and tested a simple method
how to compare a 3-dimensional distribution of proper elements.

For the Eos family, it is possible to explain its shape in the
$(a_{\rm p}, e_{\rm p}, \sin I_{\rm p})$ space
and estimate the age at the same time,
but this estimate still scales with the bulk density~$\rho$,
because most of the perturbations are non-gravitational
(including all systematic drifts ${\rm d}a/{\rm d}t$, ${\rm d}e/{\rm d}t$, ${\rm d}I/{\rm d}t$).

While we believe our model includes the key contributions,
no dynamical model is complete. For example, we miss
inner planets,
gravitational perturbations by large asteroids, or
short-term spin axis evolution due to gravitational (solar) torques.
Initial condition might be also too simple.
In particular, the velocity field might have been non-isotropic
even though in catastrophic disruptions (like Eos)
we rather expect a high degree of isotropy
\citep{Sevecek_etal_2017Icar..296..239S}.
Generally, it is better to keep both as simple as possible
to have the lowest possible number of free parameters.

Let us finally compare our nominal best-fit model to another
two distributions (size and spin) and the respective models
(collisional and rotational).


\subsection{Collisional evolution}\label{sec:collisional}

In a Monte-Carlo collisional model, size-frequency distributions
are evolved due to fragmentation and reaccumulation.
We assume two populations: the main belt, and the Eos family.
Their physical properties are summarized by the scaling law~$Q^\star_{\rm D}(r)$,
for which we assume parameters of basalt at $5\,{\rm km}\,{\rm s}^{-1}$
from \cite{Benz_Asphaug_1999Icar..142....5B}.
To compute the actual evolution, we use the Boulder code by
\cite{Morbidelli_etal_2009Icar..204..558M}.
Parametric relations in the Boulder code, which are needed to compute
the fragment distributions, are derived from SPH simulations of 
\cite{Durda_etal_2007Icar..186..498D}.

We assume the initial SFD of the main belt relatively similar to the
currently observed SFD, because we focus on the already stable
solar system, with the fixed intrinsic impact probability
$P_{\rm imp} = 3.1\times 10^{-18}\,{\rm km}^{-2}\,{\rm yr}^{-1}$
and the mean velocity $v_{\rm imp} = 5.28\,{\rm km}\,{\rm s}^{-1}$.
The initial SFD of the Eos family has the same slope
as the observed SFD in the range $D \in (15; 50)\,{\rm km}$,
and it is prolonged down to $D_{\rm min} = 0.005\,{\rm km}$.
We also account for the size-dependent dynamical decay
due to the Yarkovsky effect, with $N(t+\Delta t) = N(t)\exp(-\Delta t/\tau)$,
where the time scale~$\tau(D)$ is taken from \cite{Bottke_etal_2005Icar..179...63B}.

\begin{figure}
\centering
\begin{tabular}{c}
$t = 1300\,{\rm Myr}$ \\[-3pt]
\includegraphics[width=8cm]{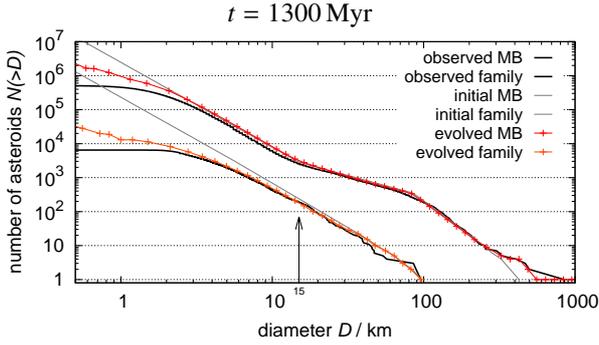} \\
\end{tabular}
\caption{The cumulative size-frequency distributions
computed by our Monte-Carlo collisional model of the two populations:
the main belt (red),
the Eos family (orange),
together with the respective initial conditions (gray),
and observations (black).
At the time around $t = 1300\,{\rm Myr}$ the correspondence is good,
except the tail below $D \lesssim 2\,{\rm km}$
where an observational incompleteness makes the SFD's shallow.
In particular, we successfully fit the knee of the family at $D \simeq 15\,{\rm km}$.}
\label{boulder_Eos_BOTTKE_DT1e7_sfd_1300}
\end{figure}

The resulting collisional evolution is shown in
Figure~\ref{boulder_Eos_BOTTKE_DT1e7_sfd_1300}.
The observed knee at $D \simeq 15\,{\rm km}$
is very important, because it usually arises from a collisional grinding.
If we start with the constant slope from above,
we can match the observed SFD at about $1.3\,{\rm Gyr}$
which is in accord with the dynamics.

It is worth to note the scaling of the age with the bulk density~$\rho$ is different
from dynamics, which in principle allows to resolve the problem.
However, the collisional model is sensitive to the initial conditions
and using a steeper SFD would result in longer age.
In other words, everything is based on the simple
assumption of the constant slope.
It would be useful to base the initial conditions
on a specific SPH model for the Eos family,
with the parent body size reaching up to 380\,{\rm km}
(according to an extrapolation of \citealt{Durda_etal_2007Icar..186..498D} results).

\subsection{Spin distribution}

\begin{figure}
\centering
\begin{tabular}{c}
$t = 0\,{\rm Myr}$ \\
\includegraphics[width=7.0cm]{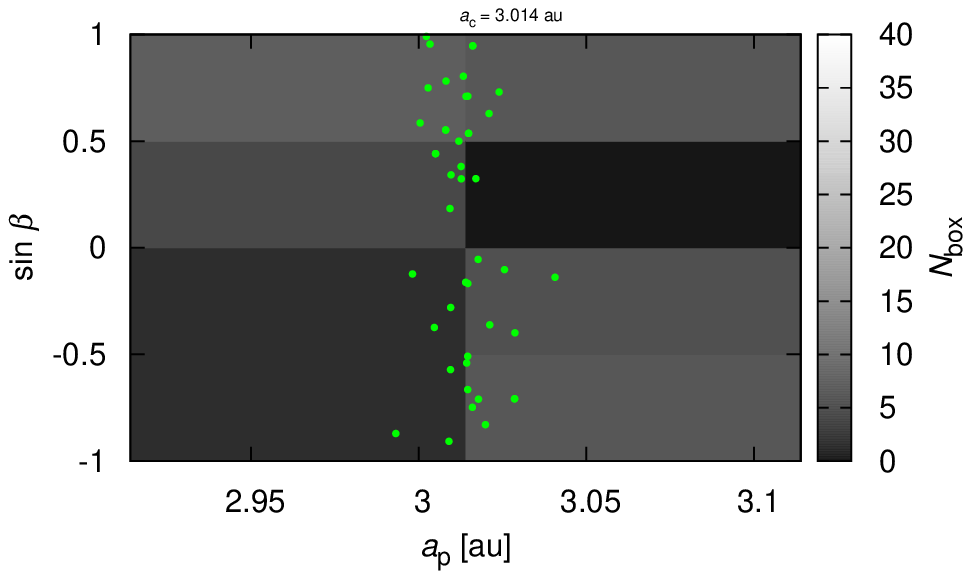} \\
$t = 1260\,{\rm Myr}$ \\
\includegraphics[width=7.0cm]{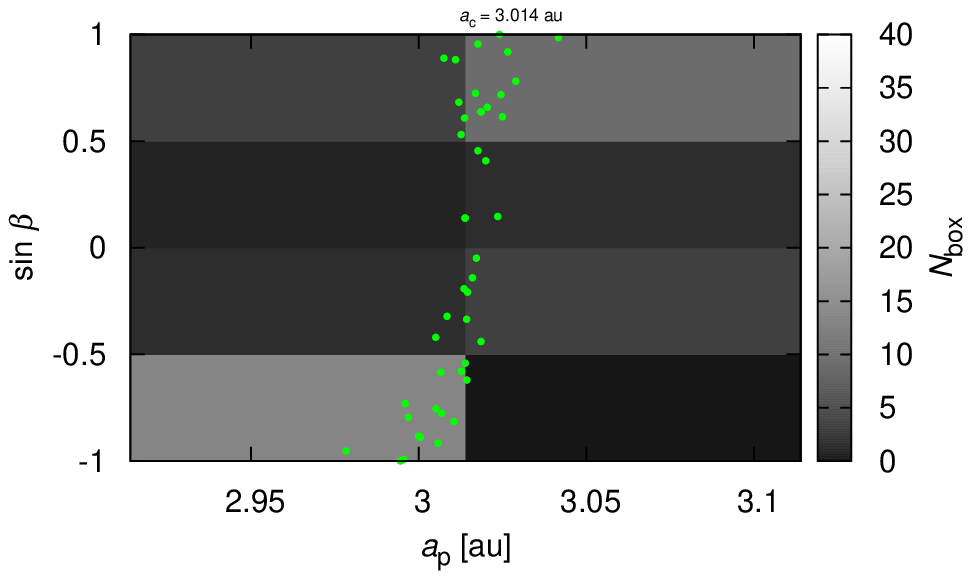} \\
observed \\
\includegraphics[width=7.0cm]{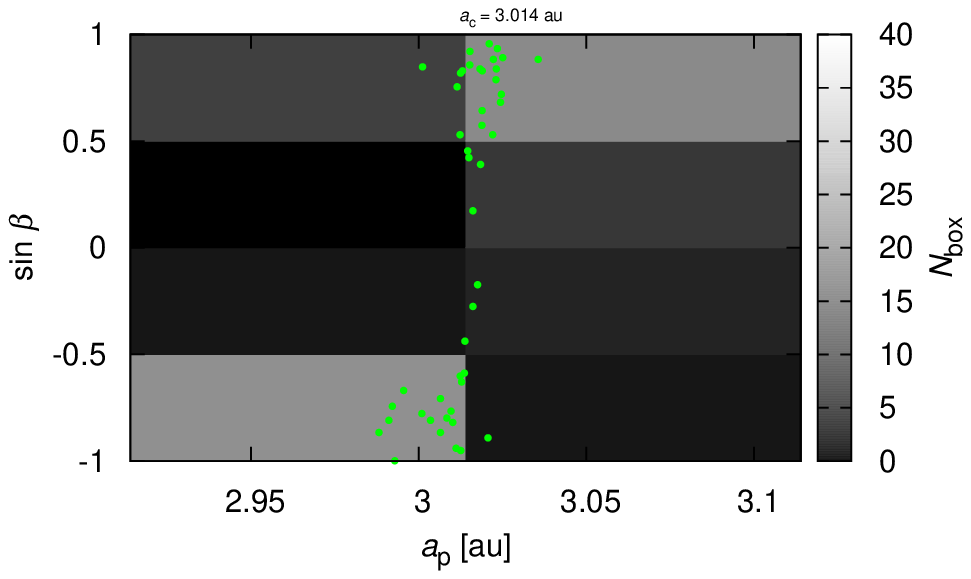} \\
\end{tabular}
\caption{The proper semimajor axis~$a_{\rm p}$ vs the sine of pole latitude~$\sin\beta$
for the initial synthetic population (top panel),
the evolved synthetic population (middle),
and the observed population of 46 bodies (bottom).
The individual bodies are shown as green dots,
while their numbers in 8~boxes are indicated by the gray scale.
The simulation started from initially isotropic random distribution,
i.e. isotropic in $\sin\beta$.
The synthetic SFD was again scaled to the observed one.
We account neither for the observational selection bias,
nor for the bias of the inversion method.
Nevertheless, it is possible to fit both
the asymmetry of the distribution with respect to $a_{\rm c} = 3.014\,{\rm au}$,
and the substantially lower number of bodies at mid-latitudes $|\sin\beta\,| < 0.5$.}
\label{eos-5_REORIENT2_abeta_syn_1260}
\end{figure}

At the same time, it is worth to check the observed
distribution of pole latitudes $\beta$,
reported in \cite{Hanus_etal_2018Icar..299...84H}.
Our dynamical model evolves the spin $(\omega, \gamma)$,
which affects the Yarkovsky drift rate ${\rm d}a/{\rm d}t$,
but we do not account for spin-orbital resonances
(so we would not explain a clustering in the Koronis family; \citealt{Slivan_2002Natur.419...49S}).
Nevertheless, if we use the current model for Eos,
with the same post-processing,
but focus on $(a_{\rm p}, \sin\beta)$ boxes instead,
we obtain the results summarized in Figure~\ref{eos-5_REORIENT2_abeta_syn_1260}.

We start from an isotropic distribution of spins,
which means isotropic also in $\sin\beta$.
After about $1.3\,{\rm Gyr}$, it is possible to fit both
the asymmetry of the distribution with respect to $a_{\rm c} = 3.014\,{\rm au}$,
and the substantially lower number of bodies at mid-latitudes $|\sin\beta\,| < 0.5$.
There are two systematics still present in our analysis,
as we account neither for the observational selection bias,
nor for the bias of the inversion method, but they should
not overturn our conclusions.

Unfortunately, the uncertainty is larger than in the nominal model,
because the number of bodies with known latitudes is limited,
namely 46 within the family core.
As a~solution, we may use the distribution of $|\beta|$
of \cite{Cibulkova_etal_2016A&A...596A..57C} which is available
for many more asteroids, but we would need to determine
the 'point-spread function', describing a~relation between
input~$|\beta|$ and output~$|\beta|$ for this (approximate) method,
which smears the distribution substantially.
Their sample also contains a~lot of bodies smaller than we had
in the previous simulations, so we would have to compute everything again.
This is postponed as a future work.

\section*{Acknowledgements}\label{sec:acknowledgements}

The work of MB has been supported by the Grant Agency of the Czech
Republic (grant no.\ P209-18-04514J). In this paper, we used observations
made by BlueEye 600 robotic observatory, supported by the Technology
Agency of the Czech Republic (grant no.\ TA03011171).
We thank the referees V.~Carruba and F.~Roig for their valuable input.


\bibliographystyle{elsarticle-harv}
\bibliography{references}

\begin{thebibliography}{35}
\expandafter\ifx\csname natexlab\endcsname\relax\def\natexlab#1{#1}\fi
\providecommand{\url}[1]{\texttt{#1}}
\providecommand{\href}[2]{#2}
\providecommand{\path}[1]{#1}
\providecommand{\DOIprefix}{doi:}
\providecommand{\ArXivprefix}{arXiv:}
\providecommand{\URLprefix}{URL: }
\providecommand{\Pubmedprefix}{pmid:}
\providecommand{\doi}[1]{\href{http://dx.doi.org/#1}{\path{#1}}}
\providecommand{\Pubmed}[1]{\href{pmid:#1}{\path{#1}}}
\providecommand{\bibinfo}[2]{#2}
\ifx\xfnm\relax \def\xfnm[#1]{\unskip,\space#1}\fi
\bibitem[{{Anderson} et~al.(1991){Anderson}, {Isaak} and
  {Oda}}]{Anderson_etal_1991JGR....9618037A}
\bibinfo{author}{{Anderson}, O.L.}, \bibinfo{author}{{Isaak}, D.L.},
  \bibinfo{author}{{Oda}, H.}, \bibinfo{year}{1991}.
\newblock \bibinfo{title}{{Thermoelastic parameters for six minerals at high
  temperature}}.
\newblock \bibinfo{journal}{\jgr} \bibinfo{volume}{96}, \bibinfo{pages}{18037}.
\newblock \DOIprefix\doi{10.1029/91JB01579}.
\bibitem[{{Benz} and {Asphaug}(1999)}]{Benz_Asphaug_1999Icar..142....5B}
\bibinfo{author}{{Benz}, W.}, \bibinfo{author}{{Asphaug}, E.},
  \bibinfo{year}{1999}.
\newblock \bibinfo{title}{{Catastrophic Disruptions Revisited}}.
\newblock \bibinfo{journal}{\icarus} \bibinfo{volume}{142},
  \bibinfo{pages}{5--20}.
\newblock \DOIprefix\doi{10.1006/icar.1999.6204},
  \href{http://arxiv.org/abs/arXiv:astro-ph/9907117}{{\tt
  arXiv:arXiv:astro-ph/9907117}}.
\bibitem[{{Bottke} et~al.(2005){Bottke}, {Durda}, {Nesvorn{\'y}}, {Jedicke},
  {Morbidelli}, {Vokrouhlick{\'y}} and
  {Levison}}]{Bottke_etal_2005Icar..179...63B}
\bibinfo{author}{{Bottke}, W.F.}, \bibinfo{author}{{Durda}, D.D.},
  \bibinfo{author}{{Nesvorn{\'y}}, D.}, \bibinfo{author}{{Jedicke}, R.},
  \bibinfo{author}{{Morbidelli}, A.}, \bibinfo{author}{{Vokrouhlick{\'y}}, D.},
  \bibinfo{author}{{Levison}, H.F.}, \bibinfo{year}{2005}.
\newblock \bibinfo{title}{{Linking the collisional history of the main asteroid
  belt to its dynamical excitation and depletion}}.
\newblock \bibinfo{journal}{\icarus} \bibinfo{volume}{179},
  \bibinfo{pages}{63--94}.
\newblock \DOIprefix\doi{10.1016/j.icarus.2005.05.017}.
\bibitem[{{Bro{\v z}} and
  {Morbidelli}(2013)}]{Broz_Morbidelli_2013Icar..223..844B}
\bibinfo{author}{{Bro{\v z}}, M.}, \bibinfo{author}{{Morbidelli}, A.},
  \bibinfo{year}{2013}.
\newblock \bibinfo{title}{{The Eos family halo}}.
\newblock \bibinfo{journal}{\icarus} \bibinfo{volume}{223},
  \bibinfo{pages}{844--849}.
\newblock \DOIprefix\doi{10.1016/j.icarus.2013.02.002},
  \href{http://arxiv.org/abs/1302.1447}{{\tt arXiv:1302.1447}}.
\bibitem[{{Bro{\v z}} et~al.(2011){Bro{\v z}}, {Vokrouhlick{\'y}},
  {Morbidelli}, {Nesvorn{\'y}} and {Bottke}}]{Broz_etal_2011MNRAS.414.2716B}
\bibinfo{author}{{Bro{\v z}}, M.}, \bibinfo{author}{{Vokrouhlick{\'y}}, D.},
  \bibinfo{author}{{Morbidelli}, A.}, \bibinfo{author}{{Nesvorn{\'y}}, D.},
  \bibinfo{author}{{Bottke}, W.F.}, \bibinfo{year}{2011}.
\newblock \bibinfo{title}{{Did the Hilda collisional family form during the
  late heavy bombardment?}}
\newblock \bibinfo{journal}{\mnras} \bibinfo{volume}{414},
  \bibinfo{pages}{2716--2727}.
\newblock \DOIprefix\doi{10.1111/j.1365-2966.2011.18587.x},
  \href{http://arxiv.org/abs/1109.1114}{{\tt arXiv:1109.1114}}.
\bibitem[{{\v{C}apek} and
  {Vokrouhlick{\'y}}(2004)}]{Capek_Vokrouhlicky_2004Icar..172..526C}
\bibinfo{author}{{\v{C}apek}, D.}, \bibinfo{author}{{Vokrouhlick{\'y}}, D.},
  \bibinfo{year}{2004}.
\newblock \bibinfo{title}{{The YORP effect with finite thermal conductivity}}.
\newblock \bibinfo{journal}{\icarus} \bibinfo{volume}{172},
  \bibinfo{pages}{526--536}.
\newblock \DOIprefix\doi{10.1016/j.icarus.2004.07.003}.
\bibitem[{{Carruba} et~al.(2013){Carruba}, {Domingos}, {Nesvorn{\'y}}, {Roig},
  {Huaman} and {Souami}}]{Carruba_etal_2013MNRAS.433.2075C}
\bibinfo{author}{{Carruba}, V.}, \bibinfo{author}{{Domingos}, R.C.},
  \bibinfo{author}{{Nesvorn{\'y}}, D.}, \bibinfo{author}{{Roig}, F.},
  \bibinfo{author}{{Huaman}, M.E.}, \bibinfo{author}{{Souami}, D.},
  \bibinfo{year}{2013}.
\newblock \bibinfo{title}{{A multidomain approach to asteroid families'
  identification}}.
\newblock \bibinfo{journal}{\mnras} \bibinfo{volume}{433},
  \bibinfo{pages}{2075--2096}.
\newblock \DOIprefix\doi{10.1093/mnras/stt884},
  \href{http://arxiv.org/abs/1305.4847}{{\tt arXiv:1305.4847}}.
\bibitem[{{Carruba} et~al.(2016){Carruba}, {Nesvorn{\'y}} and
  {Aljbaae}}]{Carruba_etal_2016Icar..271...57C}
\bibinfo{author}{{Carruba}, V.}, \bibinfo{author}{{Nesvorn{\'y}}, D.},
  \bibinfo{author}{{Aljbaae}, S.}, \bibinfo{year}{2016}.
\newblock \bibinfo{title}{{Characterizing the original ejection velocity field
  of the Koronis family}}.
\newblock \bibinfo{journal}{\icarus} \bibinfo{volume}{271},
  \bibinfo{pages}{57--66}.
\newblock \DOIprefix\doi{10.1016/j.icarus.2016.01.006},
  \href{http://arxiv.org/abs/1602.04491}{{\tt arXiv:1602.04491}}.
\bibitem[{{Carruba} et~al.(2015){Carruba}, {Nesvorn{\'y}}, {Aljbaae} and
  {Huaman}}]{Carruba_etal_2015MNRAS.451..244C}
\bibinfo{author}{{Carruba}, V.}, \bibinfo{author}{{Nesvorn{\'y}}, D.},
  \bibinfo{author}{{Aljbaae}, S.}, \bibinfo{author}{{Huaman}, M.E.},
  \bibinfo{year}{2015}.
\newblock \bibinfo{title}{{Dynamical evolution of the Cybele asteroids}}.
\newblock \bibinfo{journal}{\mnras} \bibinfo{volume}{451},
  \bibinfo{pages}{244--256}.
\newblock \DOIprefix\doi{10.1093/mnras/stv997},
  \href{http://arxiv.org/abs/1505.03745}{{\tt arXiv:1505.03745}}.
\bibitem[{{Cibulkov{\'a}} et~al.(2016){Cibulkov{\'a}}, {{\v D}urech},
  {Vokrouhlick{\'y}}, {Kaasalainen} and
  {Oszkiewicz}}]{Cibulkova_etal_2016A&A...596A..57C}
\bibinfo{author}{{Cibulkov{\'a}}, H.}, \bibinfo{author}{{{\v D}urech}, J.},
  \bibinfo{author}{{Vokrouhlick{\'y}}, D.}, \bibinfo{author}{{Kaasalainen},
  M.}, \bibinfo{author}{{Oszkiewicz}, D.A.}, \bibinfo{year}{2016}.
\newblock \bibinfo{title}{{Distribution of spin-axes longitudes and shape
  elongations of main-belt asteroids}}.
\newblock \bibinfo{journal}{\aap} \bibinfo{volume}{596}, \bibinfo{pages}{A57}.
\newblock \DOIprefix\doi{10.1051/0004-6361/201629192},
  \href{http://arxiv.org/abs/1610.02790}{{\tt arXiv:1610.02790}}.
\bibitem[{{Delbo} et~al.(2015){Delbo}, {Mueller}, {Emery}, {Rozitis} and
  {Capria}}]{Delbo_etal_2015aste.book..107D}
\bibinfo{author}{{Delbo}, M.}, \bibinfo{author}{{Mueller}, M.},
  \bibinfo{author}{{Emery}, J.P.}, \bibinfo{author}{{Rozitis}, B.},
  \bibinfo{author}{{Capria}, M.T.}, \bibinfo{year}{2015}.
\newblock \bibinfo{title}{{Asteroid Thermophysical Modeling}}.
\newblock pp. \bibinfo{pages}{107--128}.
\bibitem[{{Durda} et~al.(2007){Durda}, {Bottke}, {Nesvorn{\'y}}, {Enke},
  {Merline}, {Asphaug} and {Richardson}}]{Durda_etal_2007Icar..186..498D}
\bibinfo{author}{{Durda}, D.D.}, \bibinfo{author}{{Bottke}, W.F.},
  \bibinfo{author}{{Nesvorn{\'y}}, D.}, \bibinfo{author}{{Enke}, B.L.},
  \bibinfo{author}{{Merline}, W.J.}, \bibinfo{author}{{Asphaug}, E.},
  \bibinfo{author}{{Richardson}, D.C.}, \bibinfo{year}{2007}.
\newblock \bibinfo{title}{{Size-frequency distributions of fragments from SPH/
  N-body simulations of asteroid impacts: Comparison with observed asteroid
  families}}.
\newblock \bibinfo{journal}{\icarus} \bibinfo{volume}{186},
  \bibinfo{pages}{498--516}.
\newblock \DOIprefix\doi{10.1016/j.icarus.2006.09.013}.
\bibitem[{{Farinella} et~al.(1998){Farinella}, {Vokrouhlick{\'y}} and
  {Hartmann}}]{Farinella_etal_1998Icar..132..378F}
\bibinfo{author}{{Farinella}, P.}, \bibinfo{author}{{Vokrouhlick{\'y}}, D.},
  \bibinfo{author}{{Hartmann}, W.K.}, \bibinfo{year}{1998}.
\newblock \bibinfo{title}{{Meteorite Delivery via Yarkovsky Orbital Drift}}.
\newblock \bibinfo{journal}{\icarus} \bibinfo{volume}{132},
  \bibinfo{pages}{378--387}.
\newblock \DOIprefix\doi{10.1006/icar.1997.5872}.
\bibitem[{{Hanu{\v s}} et~al.(2018){Hanu{\v s}}, {Delbo'}, {Al{\'{\i}}-Lagoa},
  {Bolin}, {Jedicke}, {{\v D}urech}, {Cibulkov{\'a}}, {Pravec}, {Ku{\v
  s}nir{\'a}k}, {Behrend}, {Marchis}, {Antonini}, {Arnold}, {Audejean},
  {Bachschmidt}, {Bernasconi}, {Brunetto}, {Casulli}, {Dymock}, {Esseiva},
  {Esteban}, {Gerteis}, {de Groot}, {Gully}, {Hamanowa}, {Hamanowa}, {Krafft},
  {Lehk{\'y}}, {Manzini}, {Michelet}, {Morelle}, {Oey}, {Pilcher}, {Reignier},
  {Roy}, {Salom} and {Warner}}]{Hanus_etal_2018Icar..299...84H}
\bibinfo{author}{{Hanu{\v s}}, J.}, \bibinfo{author}{{Delbo'}, M.},
  \bibinfo{author}{{Al{\'{\i}}-Lagoa}, V.}, \bibinfo{author}{{Bolin}, B.},
  \bibinfo{author}{{Jedicke}, R.}, \bibinfo{author}{{{\v D}urech}, J.},
  \bibinfo{author}{{Cibulkov{\'a}}, H.}, \bibinfo{author}{{Pravec}, P.},
  \bibinfo{author}{{Ku{\v s}nir{\'a}k}, P.}, \bibinfo{author}{{Behrend}, R.},
  \bibinfo{author}{{Marchis}, F.}, \bibinfo{author}{{Antonini}, P.},
  \bibinfo{author}{{Arnold}, L.}, \bibinfo{author}{{Audejean}, M.},
  \bibinfo{author}{{Bachschmidt}, M.}, \bibinfo{author}{{Bernasconi}, L.},
  \bibinfo{author}{{Brunetto}, L.}, \bibinfo{author}{{Casulli}, S.},
  \bibinfo{author}{{Dymock}, R.}, \bibinfo{author}{{Esseiva}, N.},
  \bibinfo{author}{{Esteban}, M.}, \bibinfo{author}{{Gerteis}, O.},
  \bibinfo{author}{{de Groot}, H.}, \bibinfo{author}{{Gully}, H.},
  \bibinfo{author}{{Hamanowa}, H.}, \bibinfo{author}{{Hamanowa}, H.},
  \bibinfo{author}{{Krafft}, P.}, \bibinfo{author}{{Lehk{\'y}}, M.},
  \bibinfo{author}{{Manzini}, F.}, \bibinfo{author}{{Michelet}, J.},
  \bibinfo{author}{{Morelle}, E.}, \bibinfo{author}{{Oey}, J.},
  \bibinfo{author}{{Pilcher}, F.}, \bibinfo{author}{{Reignier}, F.},
  \bibinfo{author}{{Roy}, R.}, \bibinfo{author}{{Salom}, P.A.},
  \bibinfo{author}{{Warner}, B.D.}, \bibinfo{year}{2018}.
\newblock \bibinfo{title}{{Spin states of asteroids in the Eos collisional
  family}}.
\newblock \bibinfo{journal}{\icarus} \bibinfo{volume}{299},
  \bibinfo{pages}{84--96}.
\newblock \DOIprefix\doi{10.1016/j.icarus.2017.07.007},
  \href{http://arxiv.org/abs/1707.05507}{{\tt arXiv:1707.05507}}.
\bibitem[{{Hanu{\v s}} et~al.(2011){Hanu{\v s}}, {{\v D}urech}, {Bro{\v z}},
  {Warner}, {Pilcher}, {Stephens}, {Oey}, {Bernasconi}, {Casulli}, {Behrend},
  {Polishook}, {Henych}, {Lehk{\'y}}, {Yoshida} and
  {Ito}}]{Hanus_etal_2011A&A...530A.134H}
\bibinfo{author}{{Hanu{\v s}}, J.}, \bibinfo{author}{{{\v D}urech}, J.},
  \bibinfo{author}{{Bro{\v z}}, M.}, \bibinfo{author}{{Warner}, B.D.},
  \bibinfo{author}{{Pilcher}, F.}, \bibinfo{author}{{Stephens}, R.},
  \bibinfo{author}{{Oey}, J.}, \bibinfo{author}{{Bernasconi}, L.},
  \bibinfo{author}{{Casulli}, S.}, \bibinfo{author}{{Behrend}, R.},
  \bibinfo{author}{{Polishook}, D.}, \bibinfo{author}{{Henych}, T.},
  \bibinfo{author}{{Lehk{\'y}}, M.}, \bibinfo{author}{{Yoshida}, F.},
  \bibinfo{author}{{Ito}, T.}, \bibinfo{year}{2011}.
\newblock \bibinfo{title}{{A study of asteroid pole-latitude distribution based
  on an extended set of shape models derived by the lightcurve inversion
  method}}.
\newblock \bibinfo{journal}{\aap} \bibinfo{volume}{530}, \bibinfo{pages}{A134}.
\newblock \DOIprefix\doi{10.1051/0004-6361/201116738},
  \href{http://arxiv.org/abs/1104.4114}{{\tt arXiv:1104.4114}}.
\bibitem[{{Hirayama}(1918)}]{Hirayama_1918AJ.....31..185H}
\bibinfo{author}{{Hirayama}, K.}, \bibinfo{year}{1918}.
\newblock \bibinfo{title}{{Groups of asteroids probably of common origin}}.
\newblock \bibinfo{journal}{\aj} \bibinfo{volume}{31},
  \bibinfo{pages}{185--188}.
\newblock \DOIprefix\doi{10.1086/104299}.
\bibitem[{{Kne{\v z}evi{\'c}} and
  {Milani}(2003)}]{Knezevic_Milani_2003A&A...403.1165K}
\bibinfo{author}{{Kne{\v z}evi{\'c}}, Z.}, \bibinfo{author}{{Milani}, A.},
  \bibinfo{year}{2003}.
\newblock \bibinfo{title}{{Proper element catalogs and asteroid families}}.
\newblock \bibinfo{journal}{\aap} \bibinfo{volume}{403},
  \bibinfo{pages}{1165--1173}.
\newblock \DOIprefix\doi{10.1051/0004-6361:20030475}.
\bibitem[{{Laskar} and {Robutel}(2001)}]{Laskar_Robutel_2001CeMDA..80...39L}
\bibinfo{author}{{Laskar}, J.}, \bibinfo{author}{{Robutel}, P.},
  \bibinfo{year}{2001}.
\newblock \bibinfo{title}{{High order symplectic integrators for perturbed
  Hamiltonian systems}}.
\newblock \bibinfo{journal}{Celestial Mechanics and Dynamical Astronomy}
  \bibinfo{volume}{80}, \bibinfo{pages}{39--62}.
\newblock \href{http://arxiv.org/abs/arXiv:astro-ph/0005074}{{\tt
  arXiv:arXiv:astro-ph/0005074}}.
\bibitem[{{Levison} and {Duncan}(1994)}]{Levison_Duncan_1994Icar..108...18L}
\bibinfo{author}{{Levison}, H.F.}, \bibinfo{author}{{Duncan}, M.J.},
  \bibinfo{year}{1994}.
\newblock \bibinfo{title}{{The long-term dynamical behavior of short-period
  comets}}.
\newblock \bibinfo{journal}{\icarus} \bibinfo{volume}{108},
  \bibinfo{pages}{18--36}.
\newblock \DOIprefix\doi{10.1006/icar.1994.1039}.
\bibitem[{{Masiero} et~al.(2011){Masiero}, {Mainzer}, {Grav}, {Bauer}, {Cutri},
  {Dailey}, {Eisenhardt}, {McMillan}, {Spahr}, {Skrutskie}, {Tholen}, {Walker},
  {Wright}, {DeBaun}, {Elsbury}, {Gautier}, {Gomillion} and
  {Wilkins}}]{Masiero_etal_2011ApJ...741...68M}
\bibinfo{author}{{Masiero}, J.R.}, \bibinfo{author}{{Mainzer}, A.K.},
  \bibinfo{author}{{Grav}, T.}, \bibinfo{author}{{Bauer}, J.M.},
  \bibinfo{author}{{Cutri}, R.M.}, \bibinfo{author}{{Dailey}, J.},
  \bibinfo{author}{{Eisenhardt}, P.R.M.}, \bibinfo{author}{{McMillan}, R.S.},
  \bibinfo{author}{{Spahr}, T.B.}, \bibinfo{author}{{Skrutskie}, M.F.},
  \bibinfo{author}{{Tholen}, D.}, \bibinfo{author}{{Walker}, R.G.},
  \bibinfo{author}{{Wright}, E.L.}, \bibinfo{author}{{DeBaun}, E.},
  \bibinfo{author}{{Elsbury}, D.}, \bibinfo{author}{{Gautier}, IV, T.},
  \bibinfo{author}{{Gomillion}, S.}, \bibinfo{author}{{Wilkins}, A.},
  \bibinfo{year}{2011}.
\newblock \bibinfo{title}{{Main Belt Asteroids with WISE/NEOWISE. I.
  Preliminary Albedos and Diameters}}.
\newblock \bibinfo{journal}{\apj} \bibinfo{volume}{741}, \bibinfo{pages}{68}.
\newblock \DOIprefix\doi{10.1088/0004-637X/741/2/68},
  \href{http://arxiv.org/abs/1109.4096}{{\tt arXiv:1109.4096}}.
\bibitem[{{Morbidelli} et~al.(2009){Morbidelli}, {Bottke}, {Nesvorn{\'y}} and
  {Levison}}]{Morbidelli_etal_2009Icar..204..558M}
\bibinfo{author}{{Morbidelli}, A.}, \bibinfo{author}{{Bottke}, W.F.},
  \bibinfo{author}{{Nesvorn{\'y}}, D.}, \bibinfo{author}{{Levison}, H.F.},
  \bibinfo{year}{2009}.
\newblock \bibinfo{title}{{Asteroids were born big}}.
\newblock \bibinfo{journal}{\icarus} \bibinfo{volume}{204},
  \bibinfo{pages}{558--573}.
\newblock \DOIprefix\doi{10.1016/j.icarus.2009.07.011},
  \href{http://arxiv.org/abs/0907.2512}{{\tt arXiv:0907.2512}}.
\bibitem[{{Moth{\'e}-Diniz} et~al.(2008){Moth{\'e}-Diniz}, {Carvano}, {Bus},
  {Duffard} and {Burbine}}]{MotheDiniz_etal_2008Icar..195..277M}
\bibinfo{author}{{Moth{\'e}-Diniz}, T.}, \bibinfo{author}{{Carvano}, J.M.},
  \bibinfo{author}{{Bus}, S.J.}, \bibinfo{author}{{Duffard}, R.},
  \bibinfo{author}{{Burbine}, T.H.}, \bibinfo{year}{2008}.
\newblock \bibinfo{title}{{Mineralogical analysis of the Eos family from
  near-infrared spectra}}.
\newblock \bibinfo{journal}{\icarus} \bibinfo{volume}{195},
  \bibinfo{pages}{277--294}.
\newblock \DOIprefix\doi{10.1016/j.icarus.2007.12.005}.
\bibitem[{{Nesvorn{\'y}} et~al.(2015){Nesvorn{\'y}}, {Bro{\v z}} and
  {Carruba}}]{Nesvorny_etal_2015aste.book..297N}
\bibinfo{author}{{Nesvorn{\'y}}, D.}, \bibinfo{author}{{Bro{\v z}}, M.},
  \bibinfo{author}{{Carruba}, V.}, \bibinfo{year}{2015}.
\newblock \bibinfo{title}{{Identification and Dynamical Properties of Asteroid
  Families}}.
\newblock pp. \bibinfo{pages}{297--321}.
\bibitem[{{Novakovi\'c} and
  {Tsirvoulis}(2014)}]{Novakovic_Tsirvoulis_2014acm..conf..388N}
\bibinfo{author}{{Novakovi\'c}, B.}, \bibinfo{author}{{Tsirvoulis}, G.},
  \bibinfo{year}{2014}.
\newblock \bibinfo{title}{{Recent disruption of an asteroid from the Eos
  family}}, in: \bibinfo{editor}{{Muinonen}, K.},
  \bibinfo{editor}{{Penttil{\"a}}, A.}, \bibinfo{editor}{{Granvik}, M.},
  \bibinfo{editor}{{Virkki}, A.}, \bibinfo{editor}{{Fedorets}, G.},
  \bibinfo{editor}{{Wilkman}, O.}, \bibinfo{editor}{{Kohout}, T.} (Eds.),
  \bibinfo{booktitle}{Asteroids, Comets, Meteors 2014}.
\bibitem[{{Parker} et~al.(2008){Parker}, {Ivezi{\'c}}, {Juri{\'c}}, {Lupton},
  {Sekora} and {Kowalski}}]{Parker_etal_2008Icar..198..138P}
\bibinfo{author}{{Parker}, A.}, \bibinfo{author}{{Ivezi{\'c}}, {\v Z}.},
  \bibinfo{author}{{Juri{\'c}}, M.}, \bibinfo{author}{{Lupton}, R.},
  \bibinfo{author}{{Sekora}, M.D.}, \bibinfo{author}{{Kowalski}, A.},
  \bibinfo{year}{2008}.
\newblock \bibinfo{title}{{The size distributions of asteroid families in the
  SDSS Moving Object Catalog 4}}.
\newblock \bibinfo{journal}{\icarus} \bibinfo{volume}{198},
  \bibinfo{pages}{138--155}.
\newblock \DOIprefix\doi{10.1016/j.icarus.2008.07.002},
  \href{http://arxiv.org/abs/0807.3762}{{\tt arXiv:0807.3762}}.
\bibitem[{{Pravec} and {Harris}(2000)}]{Pravec_Harris_2000Icar..148...12P}
\bibinfo{author}{{Pravec}, P.}, \bibinfo{author}{{Harris}, A.W.},
  \bibinfo{year}{2000}.
\newblock \bibinfo{title}{{Fast and Slow Rotation of Asteroids}}.
\newblock \bibinfo{journal}{\icarus} \bibinfo{volume}{148},
  \bibinfo{pages}{12--20}.
\newblock \DOIprefix\doi{10.1006/icar.2000.6482}.
\bibitem[{{Quinn} et~al.(1991){Quinn}, {Tremaine} and
  {Duncan}}]{Quinn_etal_1991AJ....101.2287Q}
\bibinfo{author}{{Quinn}, T.R.}, \bibinfo{author}{{Tremaine}, S.},
  \bibinfo{author}{{Duncan}, M.}, \bibinfo{year}{1991}.
\newblock \bibinfo{title}{{A three million year integration of the earth's
  orbit}}.
\newblock \bibinfo{journal}{\aj} \bibinfo{volume}{101},
  \bibinfo{pages}{2287--2305}.
\newblock \DOIprefix\doi{10.1086/115850}.
\bibitem[{{\v{S}eve\v{c}ek} et~al.(2017){\v{S}eve\v{c}ek}, {Bro{\v z}},
  {Nesvorn{\'y}}, {Enke}, {Durda}, {Walsh} and
  {Richardson}}]{Sevecek_etal_2017Icar..296..239S}
\bibinfo{author}{{\v{S}eve\v{c}ek}, P.}, \bibinfo{author}{{Bro{\v z}}, M.},
  \bibinfo{author}{{Nesvorn{\'y}}, D.}, \bibinfo{author}{{Enke}, B.},
  \bibinfo{author}{{Durda}, D.}, \bibinfo{author}{{Walsh}, K.},
  \bibinfo{author}{{Richardson}, D.C.}, \bibinfo{year}{2017}.
\newblock \bibinfo{title}{{SPH/N-Body simulations of small ($D = 10$\,km)
  asteroidal breakups and improved parametric relations for Monte-Carlo
  collisional models}}.
\newblock \bibinfo{journal}{\icarus} \bibinfo{volume}{296},
  \bibinfo{pages}{239--256}.
\newblock \DOIprefix\doi{10.1016/j.icarus.2017.06.021}.
\bibitem[{{\v{S}idlichovsk{\'y}} and
  {Nesvorn{\'y}}(1996)}]{Sidlichovsky_Nesvorny_1996CeMDA..65..137S}
\bibinfo{author}{{\v{S}idlichovsk{\'y}}, M.}, \bibinfo{author}{{Nesvorn{\'y}},
  D.}, \bibinfo{year}{1996}.
\newblock \bibinfo{title}{{Frequency modified Fourier transform and its
  applications to asteroids}}.
\newblock \bibinfo{journal}{Celestial Mechanics and Dynamical Astronomy}
  \bibinfo{volume}{65}, \bibinfo{pages}{137--148}.
\newblock \DOIprefix\doi{10.1007/BF00048443}.
\bibitem[{{Slivan}(2002)}]{Slivan_2002Natur.419...49S}
\bibinfo{author}{{Slivan}, S.M.}, \bibinfo{year}{2002}.
\newblock \bibinfo{title}{{Spin vector alignment of Koronis family asteroids}}.
\newblock \bibinfo{journal}{\nat} \bibinfo{volume}{419},
  \bibinfo{pages}{49--51}.
\newblock \DOIprefix\doi{10.1038/nature00993}.
\bibitem[{{Tsirvoulis} et~al.(2018){Tsirvoulis}, {Morbidelli}, {Delbo'} and
  {Tsiganis}}]{Tsirvoulis_etal_2018Icar..304...14T}
\bibinfo{author}{{Tsirvoulis}, G.}, \bibinfo{author}{{Morbidelli}, A.},
  \bibinfo{author}{{Delbo'}, M.}, \bibinfo{author}{{Tsiganis}, K.},
  \bibinfo{year}{2018}.
\newblock \bibinfo{title}{{Reconstructing the size distribution of the
  primordial Main Belt}}.
\newblock \bibinfo{journal}{\icarus} \bibinfo{volume}{304},
  \bibinfo{pages}{14--23}.
\bibitem[{{Vokrouhlick{\'y}}(1998)}]{Vokrouhlicky_1998A&A...335.1093V}
\bibinfo{author}{{Vokrouhlick{\'y}}, D.}, \bibinfo{year}{1998}.
\newblock \bibinfo{title}{{Diurnal Yarkovsky effect as a source of mobility of
  meter-sized asteroidal fragments. I. Linear theory}}.
\newblock \bibinfo{journal}{\aap} \bibinfo{volume}{335},
  \bibinfo{pages}{1093--1100}.
\bibitem[{{Vokrouhlick{\'y}} et~al.(2006){Vokrouhlick{\'y}}, {Bro{\v z}},
  {Morbidelli}, {Bottke}, {Nesvorn{\'y}}, {Lazzaro} and
  {Rivkin}}]{Vokrouhlicky_etal_2006Icar..182...92V}
\bibinfo{author}{{Vokrouhlick{\'y}}, D.}, \bibinfo{author}{{Bro{\v z}}, M.},
  \bibinfo{author}{{Morbidelli}, A.}, \bibinfo{author}{{Bottke}, W.F.},
  \bibinfo{author}{{Nesvorn{\'y}}, D.}, \bibinfo{author}{{Lazzaro}, D.},
  \bibinfo{author}{{Rivkin}, A.S.}, \bibinfo{year}{2006}.
\newblock \bibinfo{title}{{Yarkovsky footprints in the Eos family}}.
\newblock \bibinfo{journal}{\icarus} \bibinfo{volume}{182},
  \bibinfo{pages}{92--117}.
\newblock \DOIprefix\doi{10.1016/j.icarus.2005.12.011}.
\bibitem[{{Vokrouhlick{\'y}} and
  {Farinella}(1999)}]{Vokrouhlicky_Farinella_1999AJ....118.3049V}
\bibinfo{author}{{Vokrouhlick{\'y}}, D.}, \bibinfo{author}{{Farinella}, P.},
  \bibinfo{year}{1999}.
\newblock \bibinfo{title}{{The Yarkovsky Seasonal Effect on Asteroidal
  Fragments: A Nonlinearized Theory for Spherical Bodies}}.
\newblock \bibinfo{journal}{\aj} \bibinfo{volume}{118},
  \bibinfo{pages}{3049--3060}.
\newblock \DOIprefix\doi{10.1086/301138}.
\bibitem[{{Zappal{\`a}} et~al.(1995){Zappal{\`a}}, {Bendjoya}, {Cellino},
  {Farinella} and {Froeschl{\'e}}}]{Zappala_etal_1995Icar..116..291Z}
\bibinfo{author}{{Zappal{\`a}}, V.}, \bibinfo{author}{{Bendjoya}, P.},
  \bibinfo{author}{{Cellino}, A.}, \bibinfo{author}{{Farinella}, P.},
  \bibinfo{author}{{Froeschl{\'e}}, C.}, \bibinfo{year}{1995}.
\newblock \bibinfo{title}{{Asteroid families: Search of a 12,487-asteroid
  sample using two different clustering techniques.}}
\newblock \bibinfo{journal}{\icarus} \bibinfo{volume}{116},
  \bibinfo{pages}{291--314}.
\newblock \DOIprefix\doi{10.1006/icar.1995.1127}.

\end{thebibliography}

\end{document}